\documentclass[a4paper,11pt]{article}
\pdfoutput=1 % if your are submitting a pdflatex (i.e. if you have

\usepackage{jcappub} % for details on the use of the package, please

\usepackage[T1]{fontenc} % if needed

\title{\boldmath Spherically-symmetrical vacuum solution in Freund-Nambu scalar-tensor gravity }

\author[a,b,c,1]{Akbar Davlataliev\note{Corresponding author}}
\author[d,e,2]{Bobur Turimov}
\author[b,f,h,3]{Bobomurat Ahmedov}
\author[g,4]{Yuri Vyblyi}
\author[b,5]{Chengxun Yuan}
\author[b,6]{and Chen Zhou}

\affiliation[a]{University of Tashkent for Applied Sciences, Str. Gavhar 1, Tashkent 100149, Uzbekistan}
\affiliation[b]{School of Physics, Harbin Institute of Technology, Harbin 150001, People’s Republic of China}
\affiliation[c]{New Uzbekistan University, Movarounnahr str. 1, Tashkent 100000, Uzbekistan}
\affiliation[d]{Ulugh Beg Astronomical Institute, Astronomy St. 33, Tashkent 100052, Uzbekistan}
\affiliation[e]{Engineering school, Central Asian University, Milliy bog Str.264, Tashkent, 111221,Uzbekistan}
\affiliation[f]{Institute of Theoretical Physics, National University of Uzbekistan, Tashkent 100174, Uzbekistan}
\affiliation[h]{Institute for Advanced Studies, New Uzbekistan University, Movarounnahr str. 1, Tashkent 100000, Uzbekistan}
\affiliation[g]{B. I. Stepanov Institute of Physics, 68-2, Nezavisimosti Ave., 220072, Minsk, Belarus}

\emailAdd{akbar@astrin.uz}
\emailAdd{bturimov@astrin.uz}
\emailAdd{ahmedov@astrin.uz}
\emailAdd{vyblyi@gmail.com}
\emailAdd{yuancx@hit.edu.cn}
\emailAdd{chenzhou@hit.edu.cn}

\abstract{Scalar-tensor theories of gravity represent a natural extension of general relativity, often predicting the existence of naked singularities as alternative compact object configurations. In this work, we investigate a novel, exact solution within the Freud-Nambu scalar-tensor gravity framework, which generalizes the well-known Janis-Newman-Winnicour (JNW) naked singularity spacetime through the introduction of a new parameter $q$ coupled to a real scalar field $\varphi$ with mass $\mu$. While the metric remains identical to the JNW solution, the scalar field profile is modified, offering a parametrized deformation of this important class of solutions. We conduct a comprehensive analysis of particle dynamics in this spacetime, incorporating a direct linear coupling between the test particle and the scalar field via an additional parameter $g_s$. The effects of both parameters on key astrophysical observables are systematically explored, including the specific angular momentum, the innermost stable circular orbit (ISCO), and the radiative efficiency of accretion. Furthermore, we derive the epicyclic frequencies governing oscillatory motion and investigate their implications for the quasi-periodic oscillations (QPOs) observed in black hole binary systems. Within the framework of the epicyclic resonance (ER) model, we demonstrate how the characteristic upper and lower QPO frequencies depend sensitively on the parameters $n$, $g_s$, and $q$. To constrain the model observationally, we perform a Markov Chain Monte Carlo (MCMC) analysis using twin-peak QPO data from the microquasars XTE J1550-564 and GRS 1915+105. Our analysis yields best-fit values for the black hole masses that are consistent with previous estimates ($M = 8.99 \pm 0.55 M_{\odot}$ and $M = 13.3 \pm 1.0 M_{\odot}$, respectively) and provides the first observational constraints on the new parameters, with $q \approx 3$ and $g_s \approx 0.45$. These results demonstrate that this class of modified gravity theories leaves distinctive imprints on observable phenomena, offering a viable and testable framework for exploring strong-field gravity beyond Einstein's theory.}


\begin{document}\maketitle\flushbottom

\section{Introduction}

For a long time, scalar–tensor theories of gravity have been considered promising alternatives to the standard cosmological $\Lambda$CDM model~\cite{BransDicke1961,FujiiMaeda2003,Clifton2012}. These theories introduce scalar degrees of freedom in addition to the metric tensor of general relativity, providing a broader framework to explore gravitational phenomena~\cite{SotiriouFaraoni2010,DeFelice2010}. Within this context, a particular subclass of scalar–tensor models, where the scalar field also couples directly to matter, has found wide application in the description of compact astrophysical objects such as neutron stars and black holes~\cite{DamourEspositoFarese1993,DamourEspositoFarese1996,Harada1998}. These couplings can give rise to novel phenomena, including spontaneous scalarization, modifications to the equation of state, and observable deviations from general relativity in strong-field regimes~\cite{Doneva:2022ewd,Berti2015,Barack2019}.

Among the early formulations of scalar–tensor theories is the Freund–Nambu scalar–tensor theory, a theoretical framework introduced in the late 1960s and early 1970s by Emil Freund and Yutaka Nambu~\cite{Freund1968PR}. This theory extends general relativity by incorporating a dynamical scalar field that mediates gravitational interactions alongside the metric tensor. The introduction of this scalar field provides an additional dynamical degree of freedom, which can significantly influence the evolution of the universe and the structure of compact objects~\cite{FujiiMaeda2003,Clifton2012}.

In Freund-Nambu scalar-tensor theory, gravitational action is augmented by terms involving the scalar field and its interaction with the geometry of spacetime~\cite{Freund1968PR}. These terms typically include kinetic contributions from the scalar field, its nonminimal coupling to the Ricci scalar, and, in more general formulations, possible self-interaction or potential terms~\cite{FujiiMaeda2003,SotiriouFaraoni2010}. The field equations that govern the dynamics of spacetime and the scalar field are derived by varying the action with respect to both the metric tensor and the scalar field, yielding a set of coupled nonlinear differential equations, as is characteristic of the theories of gravity scalar tensor~\cite{Clifton2012,DeFelice2010}.

One of the key motivations for studying Freund-Nambu theory lies in its potential to address some of the shortcomings of general relativity. In particular, scalar–tensor theories offer a natural framework for modifying gravitational dynamics on cosmological scales and have long been investigated as alternatives to the standard 
$\Lambda$CDM paradigm~\cite{Clifton2012,DeFelice2010}. For example, such theories provide a possible explanation for the accelerated expansion of the universe without invoking an explicit cosmological constant. Instead, the scalar field can dynamically drive this expansion, effectively playing the role of dark energy~\cite{Copeland2006,AmendolaTsujikawa2010}. In this regard, Freund-Nambu theory, as an early representative of scalar–tensor gravity, offers a natural setting for exploring alternative cosmological models and probing the nature of dark energy from a fundamental perspective~\cite{FujiiMaeda2003}.

Over the decades, Freund-Nambu scalar-tensor theory has been explored in various theoretical and phenomenological contexts, ranging from inflationary models and cosmic evolution \cite{Freund1968PR} to black hole physics and gravitational wave generation \cite{Kanti1996, Sotiriou2012}. Despite its theoretical appeal, the empirical viability of the theory remains an open question, subject to ongoing observational scrutiny and experimental tests \citep{Will2014}.

A simple dark energy model based on the Freund-Nambu framework has been discussed in~\cite{Dudko2016GC}, where the authors analyze the cosmological implications of a dynamically evolving scalar field in late-time universe dynamics.

In the framework of general relativity, a naked singularity refers to a gravitational singularity that is not hidden behind an event horizon. In contrast to black holes, where the singular region is enclosed by an event horizon that prevents information from escaping, a naked singularity would be directly visible to distant observers. This possibility raises profound questions about the predictability and determinism of gravitational physics. Mathematically, a singularity corresponds to a region of spacetime where curvature invariants diverge, and the classical description of gravity breaks down. For example, scalars constructed from the Riemann tensor, such as the Kretschmann scalar, become unbounded at the singular point. A naked singularity arises when the singular region is not shielded by an event horizon. In such cases, null and timelike geodesics originating arbitrarily close to the singularity can reach distant observers. One of the simplest spacetimes describing a naked singularity is the Janis–Newman–Winicour (JNW) solution~\cite{Janis68}. This metric arises as an exact solution of the coupled Einstein–scalar field equations and represents a generalization of the Schwarzschild spacetime. In comparison with the Schwarzschild solution, the JNW geometry contains an additional parameter associated with the presence of a massless scalar field, which significantly modifies the spacetime structure and may lead to the formation of a naked singularity instead of an event horizon.

The motion of test particles in static spherically symmetric spacetimes provides a fundamental probe of the underlying gravity theory. By studying geodesic motion, key physical quantities can be extracted, such as the innermost stable circular orbit (ISCO), specific energy and angular momentum, and epicyclic frequencies \cite{stuchlik2013,bambi2017,Aliev:2006qi, 2026AnPhy.48670332N, Turakhonov:2024smp,Turakhonov:2024xfg, Ibrokhimov:2024hxg,Umarov:2025ihy, Umarov:2025wzm,Turimov:2024hwh,Turimov:2024tvt,Turimov:2025odi,Turimov:2025tmf,Turimov:2022iff,Boboqambarova:2021cbf,Turimov:2021jgk,Turimov:2020fme}. These quantities directly influence observable astrophysical phenomena, particularly the quasi-periodic oscillations (QPOs) observed in the X-ray flux of accreting black hole binaries~\cite{2024ChJPh..92..143R,2025PDU....5002102R}. QPOs often appear as twin peaks with characteristic frequency ratios, and the epicyclic resonance (ER) model explains these features by identifying the upper frequency $\nu_U$ with the orbital frequency $\nu_\phi$ and the lower frequency $\nu_L$ with the radial epicyclic frequency $\nu_r$ \cite{torok2005, stuchlik2008}. This framework has been successfully applied to constrain black hole parameters in several microquasars, making QPOs an invaluable diagnostic for testing alternative gravity theories in the strong-field regime \cite{stuchlik2013, kolos2023}.

To confront theoretical models with observational data, robust statistical methods for parameter estimation are essential. Markov Chain Monte Carlo (MCMC) methods have become an indispensable tool for this purpose, efficiently exploring multi-dimensional parameter spaces to construct posterior probability distributions given observed data \cite{sharma2017, foreman-mackey2013, Shabbir:2026qlh, Shermatov:2025ljg, Zahra:2025tdo,Shermatov:2025rpj,Turimov:2024orr}. In black hole astrophysics, MCMC analysis of twin-peak QPOs from sources such as XTE J1550-564 and GRS 1915+105 has been widely used to estimate black hole masses and to constrain deviations from standard spacetime metrics \cite{shafee2006, kolos2023, Hoshimov:2025tdx}. Recent studies have demonstrated that combining geodesic analysis in static spherically symmetric spacetimes with MCMC techniques can provide sensitive tests of scalar-tensor theories and other modifications of general relativity \cite{stuchlik2021, bambi2018}.

This paper is structured as follows: In Sect.~\ref{Sec:Formulations}, we present the foundational equations of the Freund-Nambu scalar-tensor theory, including the action and the corresponding field equations. In Sect.~\ref{Sec:ScalarF} and in Sect.~\ref{sec.4}, we construct and analyze a general solution to the Einstein-scalar field equations within this theoretical framework. Section~\ref{Sec.5} is devoted to the study of particle dynamics, including the derivation of constants of motion and the effective potential. In Sect.~\ref{sec.6}, we investigate oscillatory motion and derive the epicyclic frequencies governing small perturbations of circular orbits. Section~\ref{sec.7} presents the observational constraints obtained from the twin-peak QPO data using MCMC analysis. Finally, in Sect.~\ref{Sec:Conclusions}, we summarize the key results and outline possible directions for future research.

\section{Freund-Nambu scalar-tensor gravity\label{Sec:Formulations}}

The action for the Einstein–scalar field system within the framework of Freund–Nambu scalar-tensor gravity, expressed in geometrized units $G = c = 1$, is given by~\cite{Freund1968PR}
\begin{align}\label{action}
{\cal S} = \frac{1}{16\pi} \int d^4x, \sqrt{-g} \left(R - \frac{2\partial_\alpha\varphi\partial^\alpha\varphi}{1 + 2q\varphi} + 2\mu^2 \varphi^2 \right) + {\cal S}_M \ ,
\end{align}
where $g = |g_{\alpha\beta}|$ is the determinant of the spacetime metric, $R$ is the Ricci scalar, and $\varphi$ is a real scalar field with mass $\mu$. The constant $q$ governs the strength of the coupling between the scalar field and the geometry. The term ${\cal S}_M$ represents the action of matter, which is constructed from the matter fields $Q_M$ and is coupled to the conformally rescaled metric $(1 + 2q\varphi)g_{\alpha\beta}$, such that ${\cal S}_M = {\cal S}_M\left[(1 + 2q\varphi)g_{\alpha\beta},Q_M\right]$.

Varying the total action \eqref{action} with respect to the metric tensor $g_{\alpha\beta}$ and the scalar field $\varphi$, we derive the field equations governing the dynamics of the coupled Einstein–scalar system. These are given by
\begin{align}\label{eq}
&G_{\alpha\beta}=R_{\alpha\beta}-\frac{1}{2}g_{\alpha\beta}R=T_{\alpha\beta}+T^M_{\alpha\beta} \ , \\
\label{eqf}
&(\square - \mu^2)\varphi = q\left( T + T^M \right) \ ,
\end{align}
where $G_{\alpha\beta}$ is the Einstein tensor and $\square = -\nabla^\alpha\nabla_\alpha$ is the covariant D’Alembert operator. The energy-momentum tensor $T_{\alpha\beta}$ for the scalar field takes the form
\begin{align}
T_{\alpha\beta} = \frac{2\partial_\alpha\varphi\partial_\beta\varphi}{1 + 2q\varphi} - g_{\alpha\beta} \left( \frac{\partial_\gamma\varphi,\partial^\gamma\varphi}{1 + 2q\varphi} - \mu^2 \varphi^2 \right) \ ,
\end{align}
with its trace given by
\begin{align}
T = g^{\alpha\beta} T_{\alpha\beta} = -\frac{2\partial_\alpha\varphi\partial^\alpha\varphi}{1 + 2q\varphi} + 4\mu^2 \varphi^2 \ .
\end{align}

In the present work, our aim is to construct an exact analytical solution to the Einstein equations sourced by a massless scalar field in Freund-Nambu theory. To simplify the problem, we consider the vacuum configuration by neglecting both the scalar field mass term (i.e., setting $\mu = 0$) and the matter sector (i.e., setting ${\cal S}_M = 0$). This reduction allows us to focus on the purely geometric effects of a self-interacting scalar field on a curved spacetime background. In the following section, we provide a detailed derivation and analysis of the resulting field equations and the corresponding exact solutions.

\section{Massless scalar field solution\label{Sec:ScalarF}}

After ignoring the matter term, the system of equations for the Einstein-massless scalar field in the Freund-Nambu model can be expressed as
\begin{align}
\square\varphi=-\frac{q}{1+2q\varphi}\partial_\alpha\varphi\partial^\alpha\varphi\ .
\end{align}
Thus, the equation for the massless scalar field can be rewritten as follows:
\begin{align}\label{KG}
&\frac{1}{\sqrt{-g}}\partial_\mu\left(\sqrt{-g}\partial^\mu\varphi\right)=\frac{q}{1+2q\varphi}\partial_\alpha\varphi\partial^\alpha\varphi\ .
\end{align}
Assume that the scalar field is stationary and depends only on the radial coordinate, i.e., $\varphi=\varphi(r)$. So, equation (\ref{KG}) yields
\begin{align}
&\frac{1}{\sqrt{-g}}\frac{d}{dr}
\left(\sqrt{-g}g^{rr}\frac{d\varphi}{dr}\right)=\frac{q}{1+2q\varphi}g^{rr}\left(\frac{d\varphi}{dr}\right)^2\ .
\end{align}
Hereafter, introducing new field $F=1+2q\varphi$, the equation reduces to
\begin{align}
\left(\sqrt{-g}g^{rr}\frac{dF}{dr}\right)^{-1}\frac{d}{dr}
\left(\sqrt{-g}g^{rr}\frac{dF}{dr}\right)=\frac{1}{2F}\left(\frac{dF}{dr}\right)\ ,
\end{align}
which can be rewritten as 
\begin{align}
&\frac{d}{dr}\ln\left(\sqrt{-g}g^{rr}\frac{dF}{dr}\right)=\frac{d}{dr}\ln\sqrt{F}\ ,
\end{align}
and after integrating the above equation, one obtains
\begin{align}
&\frac{dF}{\sqrt{F}}=\frac{2C_1dr}{\sqrt{-g}g^{rr}}\ ,
\end{align}
or
\begin{align}\label{solF}
&\sqrt{F}=\int\frac{C_1dr}{\sqrt{-g}g^{rr}}+C_2\ ,
\end{align}
where $C_1$ and $C_2$ are constants of integration. Once the spacetime metric is known, the expression for the scalar field can be immediately derived from the expression \eqref{solF}. Later on the physical meaning of the constant will be introduced.

\section{Spherically-symmetric solutions}\label{sec.4}

Now we focus on the Einstein field equations. In the case of a massless scalar field, equation (\ref{eq}) can be rewritten as 
\begin{align}\label{EEq}
&G_{\alpha\beta}=T_{\alpha\beta}=\frac{1}{1+2q\varphi}\left(2\partial_\alpha\varphi\partial_\beta\varphi-g_{\alpha\beta}\partial_\gamma\varphi\partial^\gamma\varphi\right)\ ,
\end{align}
Since the scalar field depends on the radial coordinate, the equation (\ref{EEq}) can be expressed as 
\begin{align}\label{main}
G^0_{~~0}=-\frac{\partial_r\varphi\partial^r\varphi}{1+2q\varphi}=-G^1_{~~1}=G^2_{~~2}=G^3_{~~3}\ .    
\end{align}
Before moving on further we use the following ansatz $g_{tt}g_{rr}=-1$. So, the spacetime metric around the spherically symmetric object in FN theory can be given as
\begin{align}\label{metric}
ds^2=-e^{\nu}dt^2+e^{-\nu}\left[dr^2+\lambda(d\theta^2+\sin^2\theta d\phi^2)\right]\ ,
\end{align}
where $\nu=\nu(r)$ and $\lambda=\lambda(r)$ are unknown radial functions. The explicit form of the components of the Einstein tensor can be expressed in terms of radial profile functions in the following form:
\begin{align}\label{Gtt}
&G^0_{~~0} = \frac{1}{4}e^{\nu}\left(\frac{4\lambda''-4\lambda'\nu'-4}{\lambda}-\frac{\lambda'^2}{\lambda^2}+\nu'^2-4\nu''\right)\ ,\\\label{Grr}
&G^1_{~~1} = \frac{1}{2}e^{\nu}\left(\frac{\lambda'^2}{\lambda^2}-\nu'^2-\frac{4}{\lambda}\right)\ ,\\\label{Gqq}
&G^2_{~~2} = \frac{1}{4}e^{\nu}\left(-\frac{\lambda'^2}{\lambda^2}+\nu'^2+\frac{2\lambda''}{\lambda}\right) = G^3_{~~3}\ .
\end{align} 
Consequently, using the Einstein equations (\ref{main}), one can obtain~\cite{Turimov2022particles,Turimov2021PDU}  
\begin{align}\label{G0}
G^1_{~~1}&+G^2_{~~2}=0=\frac{e^{\nu}}{2\lambda}\left(\lambda''-2\right)\ ,
\\\label{GV}
G^0_{~~0}&+G^1_{~~1}=0=\frac{e^{\nu}}{\lambda}\left(\lambda''-\lambda\nu''-\lambda'\nu'-2\right) \ ,
\\\label{GS}
G^1_{~~1}&-G^0_{~~0}=\frac{2\partial_r\varphi\partial^r\varphi}{1+2q\varphi}=e^{\nu}\left(\frac{\lambda'\nu'-\lambda''}{\lambda}+\frac{\left(\lambda'\right)^2}{2 \lambda^2}+\nu''-\frac{\left(\nu'\right)^2}{2}\right)\ ,
\end{align}
where the prime denotes the derivatives with respect to the radial coordinate. From the above equations, simple differential equations for $\lambda$ and $\nu$ can be obtained as follows:
\begin{align}\label{eqlambda}
&\lambda''-2=0\ ,
\\\label{eqnu}
&\lambda\nu''+\lambda'\nu'=0\ .
\end{align}
The solution of (\ref{eqlambda}) is rather simple and can be written in terms of two integration constants as follows $\lambda=r^2+2C_3r+C_4$. Hereafter performing simple algebraic manipulations equation (\ref{eqnu}) yields 
\begin{align}\label{eqnuu}
\frac{\nu''}{\nu'}+\frac{\lambda'}{\lambda}=\left[\ln(\nu'\lambda)\right]'=0\ ,    
\end{align}
and solution above equation can be expressed as follows:
\begin{align}\label{solnu}
\nu=C_5\int \frac{dr}{\lambda}+C_6\ ,    
\end{align}
where $C_3$, $C_4$, $C_5$ and $C_6$ are constants of integration in two metric functions. Now we can focus on the explicit form of the solution to function $\nu$. By substituting the explicit form of the function $\lambda$ into equation \eqref{solnu}, one can obtain:
\begin{align}\nonumber
\nu&=C_5\int \frac{dr}{r^2+2C_3r+C_4}+C_6
\\\nonumber
&=C_5\int \frac{d(r+C_3)}{(r+C_3)^2-C_3^2+C_4}+C_6 
\\\nonumber
&=\frac{C_5}{2\sqrt{C_3^2-C_4}}\ln\left(\frac{r+C_3-\sqrt{C_3^2-C_4}}{r+C_3+\sqrt{C_3^2-C_4}}\right)+C_6
\\
&=\ln\left(\frac{r+C_3-\sqrt{C_3^2-C_4}}{r+C_3+\sqrt{C_3^2-C_4}}\right)^{\frac{C_5}{2\sqrt{C_3^2-C_4}}}+C_6\ .
\end{align}
At the large distance the spacetime metric \eqref{metric} reduces to Minkowski spacetime, therefore constant $C_6$ vanishes (i.e. $C_6=0$). Finally, one can obtain
\begin{align}\label{exp}
e^\nu=\left(\frac{r+C_3-\sqrt{C_3^2-C_4}}{r+C_3+\sqrt{C_3^2-C_4}}\right)^{\frac{C_5}{2\sqrt{C_3^2-C_4}}}\ .
\end{align}

Let us now calculate the scalar field in equation \eqref{solF}, using the determinant of the metric function $g=-\lambda^2 e^{-2\nu}$, it is easy to check that
\begin{align}
\sqrt{F}&=C_1\int\frac{dr}{\lambda}+C_2\ ,
\end{align}
or
\begin{align}
\sqrt{1+2q\varphi}=\frac{C_1}{2\sqrt{C_3^2-C_4}}\ln\left(\frac{r+C_3-\sqrt{C_3^2-C_4}}{r+C_3+\sqrt{C_3^2-C_4}}\right)+C_2\ .
\end{align}
One has to emphasise that at the large distance the scalar field vanishes, therefore from above equation one can find that $C_2=1$. If $q=0$ left-hand side of equation will be equal to $1$, however right-hand side will the same, therefore constant $C_1$ should be proportional to parameter $q$ (i.e. $C_1=qC$). The equation for scalar field takes a form:
\begin{align}
\sqrt{1+2q\varphi}=\frac{qC}{2\sqrt{C_3^2-C_4}}\ln\left(\frac{r+C_3-\sqrt{C_3^2-C_4}}{r+C_3+\sqrt{C_3^2-C_4}}\right)+1\ ,
\end{align}
or
\begin{align}\nonumber\label{phi}
\varphi&=\frac{C}{2\sqrt{C_3^2-C_4}}\ln\left(\frac{r+C_3-\sqrt{C_3^2-C_4}}{r+C_3+\sqrt{C_3^2-C_4}}\right)\\&\times\left[1+\frac{qC}{4\sqrt{C_3^2-C_4}}\ln\left(\frac{r+C_3-\sqrt{C_3^2-C_4}}{r+C_3+\sqrt{C_3^2-C_4}}\right)\right]\ .
\end{align}

Now using equation \eqref{main} together with \eqref{phi} and \eqref{exp}, the relation between constants can be obtained as follows
\begin{align}
C^2=4C_3^2-4C_4-C_5^2\ .    
\end{align}

After performing the following coordinate transformation $r+C_3+\sqrt{C_3^2-C_4}\to r$, the simple expressions for unknown functions can be obtained as follows:
\begin{align}
&\varphi=\frac{C}{2\sqrt{C_3^2-C_4}}\ln\left(1-\frac{2\sqrt{C_3^2-C_4}}{r}\right)\left[1+\frac{qC}{4\sqrt{C_3^2-C_4}}\ln\left(1-\frac{2\sqrt{C_3^2-C_4}}{r}\right)\right]\ ,
\\
&e^\nu=\left(1-\frac{2\sqrt{C_3^2-C_4}}{r}\right)^{\frac{C_5}{2\sqrt{C_3^2-C_4}}}\ ,
\\
&\lambda=r^2\left(1-\frac{2\sqrt{C_3^2-C_4}}{r}\right)\ .
\end{align}
The physical meaning of remaining constants are related to the mass of the central object $M$ and scalar parameter $n$ in the following form:
$$
\sqrt{C_3^2-C_4}=\frac{M}{n}\ ,\qquad \frac{C_5}{2\sqrt{C_3^2-C_4}}=n\ .
$$
Consequently, the spherical symmetric solution in Freund-Nambu theory can be expressed as
\begin{align}\label{JNW}
ds^2&=-\left(1-\frac{2M}{nr}\right)^{n}dt^2+\left(1-\frac{2M}{nr}\right)^{-n}dr^2+r^2\left(1-\frac{2M}{nr}\right)^{1-n}(d\theta^2+\sin^2\theta d\phi^2)\ ,
\end{align}
where $M$ is the mass of the object, and the parameter $n$ arises due to the scalar field. The associated scalar field is 
\begin{align}
\varphi=\frac{\sqrt{1-n^2}}{2}\ln\left(1-\frac{2M}{nr}\right)\left[1+\frac{q}{2}\frac{\sqrt{1-n^2}}{2}\ln\left(1-\frac{2M}{nr}\right)\right]\ .    
\end{align}

If we consider the Einstein-scalar field system in the framework of general relativity, the coupling constant $q$ should vanish, i.e., $q=0$. In this case, the solution (\ref{JNW}) reduces to the well-known Janis-Newman-Winicour solution~\cite{Janis68}, which will be given by spacetime in equation \eqref{JNW}, while the associated scalar field yields 
\begin{align}
\varphi=\frac{\sqrt{1-n^2}}{2}\ln\left(1-\frac{2M}{nr}\right)\ .    
\end{align}

\section{Particle dynamics\label{Sec.5}}

The action for the relativistic particle of mass $m$ in the presence of the scalar field $\varphi$ is given as~\cite{Breuer1973PRD,Turimov:2024orr,Turimov:2024hwh}
\begin{align}\label{actionEOM}
S=-\int m_* ds\ , \qquad m_*=m(1+g_s\varphi)\ ,    
\end{align}
where $m_*$ is the effective mass of the test particle in the presence of a scalar field, and $g_s$ is a coupling constant between scalar and massive particle. We aim to investigate particle dynamics in the vicinity of gravitational object described by the spacetime metric \eqref{metric}, which is the exact solution of Einstein-scalar field equations in Fread-Nambu theory. From the action \eqref{actionEOM}, equation of motion yields
\begin{align}\label{eq3}
&u^\mu\nabla_\mu u^\nu=\frac{g_s}{1+g_s\varphi}(g^{\mu\nu}+u^\mu u^\nu)\nabla_\mu\varphi\ .
\end{align}
Using the equation of motion \eqref{eq3}, the Lagrangian for a massive particle influenced by a scalar field is given by:
\begin{align}
L = \frac{1}{2}(1 + g_s \varphi) g_{\mu\nu} u^\mu u^\nu \ ,    
\end{align}
The corresponding four-momentum of the particle is
\begin{align}
P_\mu = \frac{\partial L}{\partial u^\mu} = (1 + g_s \varphi) g_{\mu\nu} u^\nu \ .    
\end{align}
Since the spacetime geometry is independent of the coordinates \( t \) and \( \phi \), the corresponding components of the four-momentum, \( P_t \) and \( P_\phi \), are conserved. These correspond to two conserved quantities: the energy and angular momentum of the particle. The specific energy \( \mathcal{E} \) and specific angular momentum \( \mathcal{L} \) can be expressed as:
\begin{align}
\mathcal{E} = -\frac{P_t}{m} = -(1 + g_s \varphi) g_{tt} u^t\ , \quad
\mathcal{L} = \frac{P_\phi}{m} = (1 + g_s \varphi) g_{\phi\phi} u^\phi \ .    
\end{align}
Applying the normalization condition for the four-velocity of the particle leads to
\begin{align}
g_{rr} \dot{r}^2 + g_{\theta\theta} \dot{\theta}^2 + 1 + \frac{1}{(1 + g_s \varphi)^2} \left( \frac{\mathcal{E}^2}{g_{tt}} + \frac{\mathcal{L}^2}{g_{\phi\phi}} \right) = 0\ . \label{eom}    
\end{align}
For simplicity, we restrict our analysis to equatorial circular motion, characterized by $\theta=\pi/2$ and $\dot{\theta}=0$. Within this setup, we investigate the properties of the innermost stable circular orbit (ISCO) of a massive test particle in the Janis–Newman–Winicour (JNW) spacetime. Particular emphasis is placed on the role of the background scalar field, including the impact of its nonlinear self-interactions on the orbital stability. This allows us to assess how scalar-field-induced deviations from the Schwarzschild geometry modify the location and stability of circular orbits in the strong-field regime. After simple algebraic manipulations, Eq.~\eqref{eom} can be rewritten as
\begin{align}\label{EOS}
(1 + g_s \varphi)^2 \dot{r}^2 = \mathcal{E}^2 - V(r)\ ,    
\end{align}
where
\begin{align}
V(r)=\left(1-\frac{2M}{nr}\right)^n(1+g_s\varphi)^2+\frac{\mathcal{L}^2}{r^2}\left(1-\frac{2M}{nr}\right)^{2n-1}\ ,    
\end{align}
is the effective potential that governs the radial motion of the particle.

The ISCO radius is a key concept in understanding dynamics near compact objects like black holes and neutron stars. It represents the smallest radius at which a test particle can maintain a stable circular orbit. Any perturbation within this radius typically results in the particle spiraling inward and eventually accreting onto the central object. Circular orbits are determined by the conditions $\dot{r}=0$ and $\partial_r V(r)=0$, which fix the conserved energy $\mathcal{E}$ and angular momentum $\mathcal{L}$ of the particle at a given orbital radius. The stability of these circular trajectories is then governed by the second derivative of the effective potential. In particular, stable circular motion requires $\partial_r^2 V(r)>0$, while the marginally stable orbit—identified with the innermost stable circular orbit (ISCO)—is defined by the simultaneous conditions
\begin{align}\label{cond}
V(r_{\rm ISCO})=\mathcal{E}^2,\qquad
\left.\frac{dV}{dr}\right|_{r={\rm ISCO}}=0,\qquad
\left.\frac{d^2V}{dr^2}\right|_{r={\rm ISCO}}=0\ .
\end{align}
The presence of the scalar field, encoded through the coupling $(1+g_s\varphi)$, modifies both the location and stability properties of circular orbits compared to the JNW limit given by~\cite{Turimov:2024oxn} 
\begin{align}
r_{\rm ISCO} = \left(3 + \frac{1}{n} + \sqrt{5 - \frac{1}{n^2}} \right)M \ ,   
\end{align}
when coupling constant is vanished $g_s=0$. However, in the presence of a nonvanishing coupling constant, obtaining exact analytical solutions of the conditions in Eq.~\eqref{cond} becomes highly nontrivial. This difficulty stems from the fact that nonlinear scalar-field contributions enter the effective potential multiplicatively, inducing nontrivial modifications to the orbital structure. As a consequence, the ISCO radius acquires a sensitive dependence on both the scalar charge and the coupling strength. Therefore, numerical methods are generally required to determine the dependence of the ISCO position on the coupling parameter $g_s$ and the nonlinearity parameter $q$. Such an analysis provides a powerful probe of deviations from general relativity in the strong-field regime, with direct implications for accretion processes and disk dynamics around compact objects described by the spacetime geometry~\eqref{metric}. In this context, it is particularly instructive to examine how the presence of the scalar field alters the ISCO structure in the JNW spacetime.

\begin{figure}
    \centering
    \includegraphics[width=0.45\linewidth]{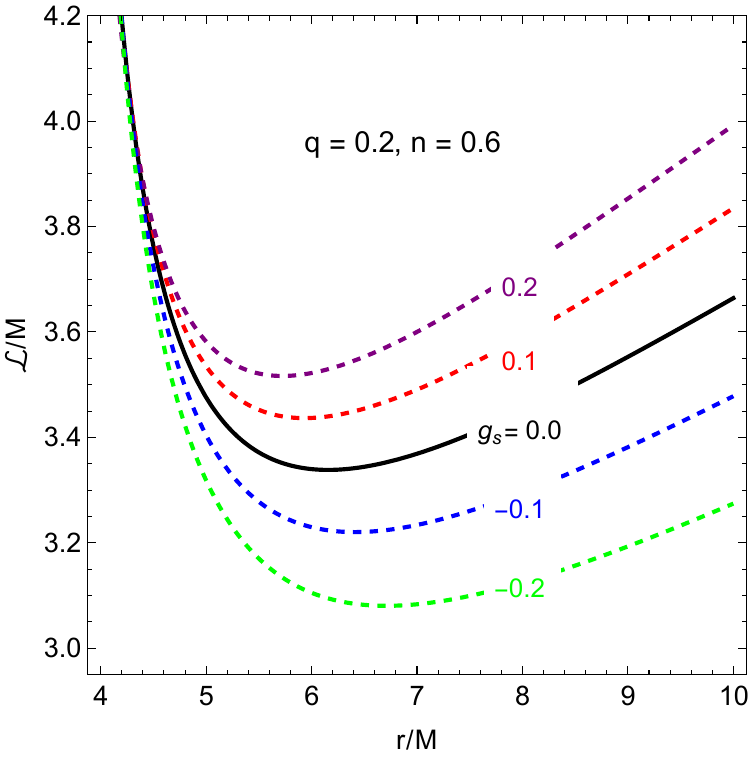}
    \includegraphics[width=0.45\linewidth]{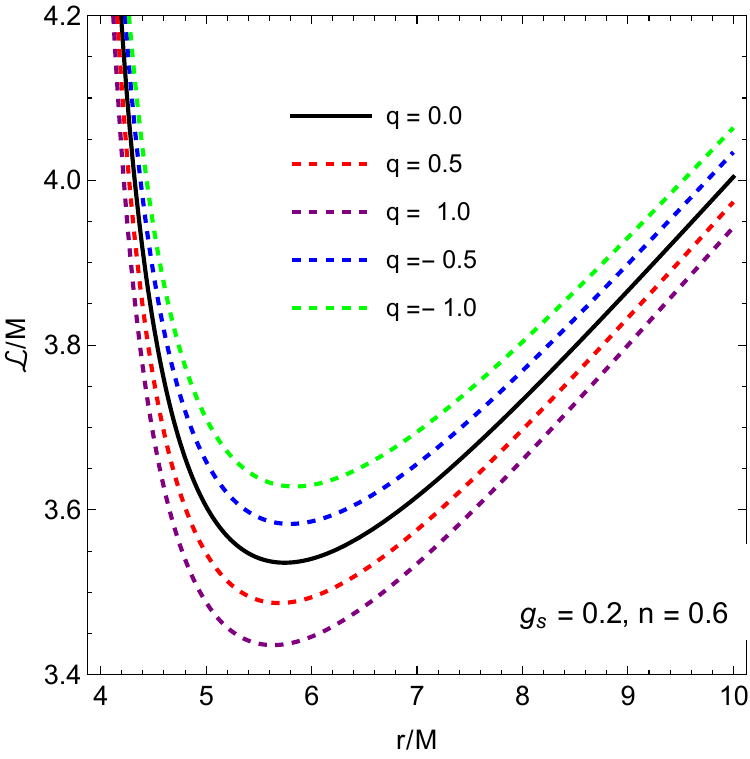}
    \caption{Radial dependencies of specific angular momentum for various values of $g_s$ and $q$ on fixed values of spacetime parameter $n$.}
    \label{figL}
\end{figure}

By imposing the first two conditions in Eq.~\eqref{cond}, namely $V(r)=\mathcal{E}^2$ and $dV/dr=0$, one obtains the critical specific energy and angular momentum of a massive particle in circular orbit as
\begin{align}
\mathcal{E}^2 &=
\left(1+ g_s\varphi\right) \left(1-\frac{2 M}{nr}\right)^n \left(1+g_s \left(r \varphi '+\varphi \right)\right)\ , \label{E2}
\\
\mathcal{L}^2 &=r^3 g_s \varphi ' \left(1+ g_s\varphi\right) \left(1-\frac{2 M}{nr}\right)^{1-n}\ . \label{L2}
\end{align}
The ISCO radius is identified by locating the stationary points of these expressions with respect to the radial coordinate $r$, corresponding to the marginal stability condition $d^2V/dr^2=0$. To illustrate this behavior, Fig.~\ref{figL} shows the radial dependence of the critical specific energy and specific angular momentum of a massive particle. The minima of these curves mark the ISCO locations for given values of the coupling and nonlinearity parameters, thereby highlighting the impact of the scalar field on orbital stability. In particular, we find that the coupling parameter shifts the ISCO radius shift due to the coupling and nonlinearity parameters, as inferred from the minima of the critical specific energy and angular momentum in Fig.~\ref{figL}.   

\begin{figure}
    \centering
    \includegraphics[width=0.45\linewidth]{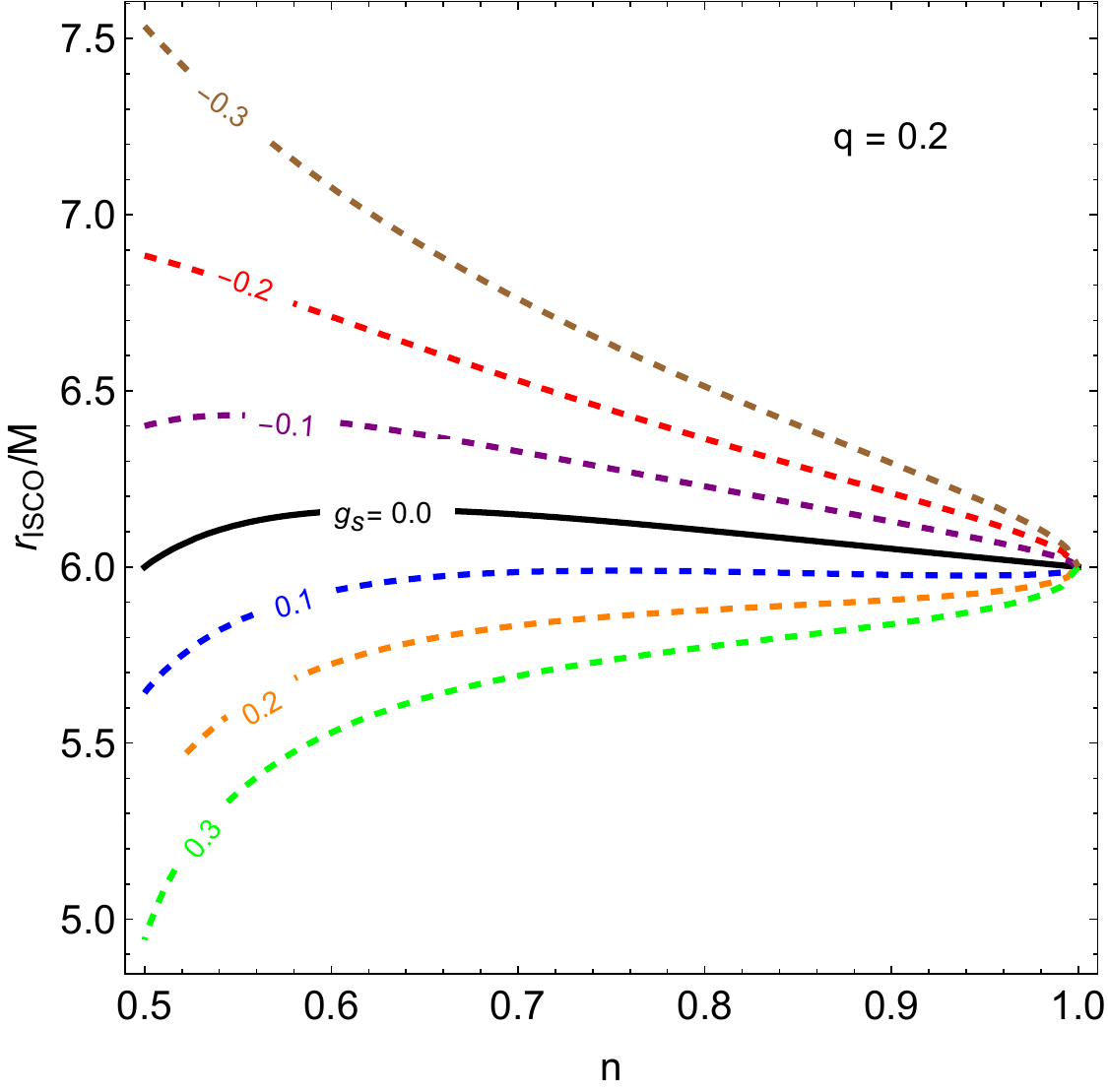}
    \includegraphics[width=0.45\linewidth]{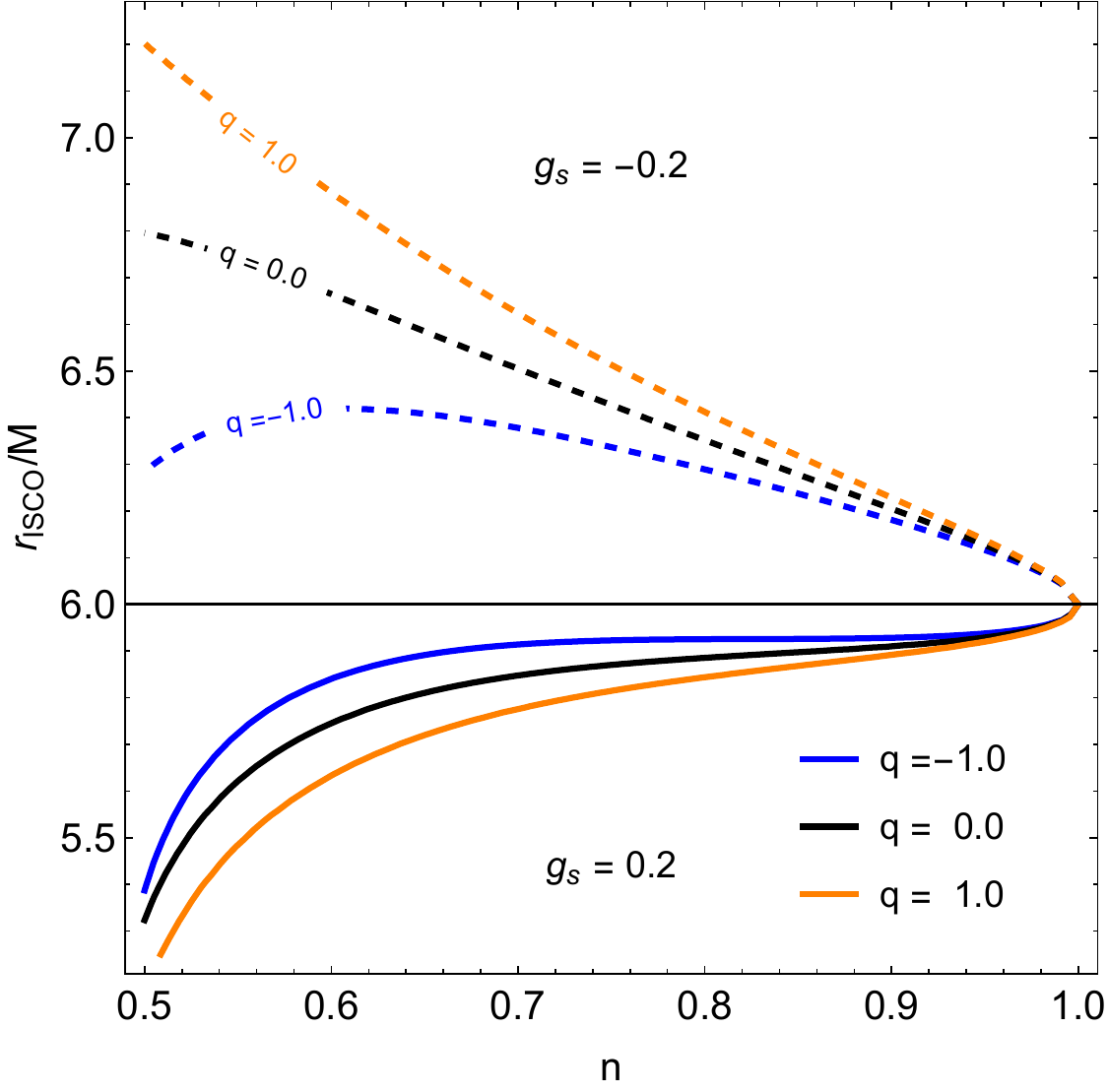}
    \caption{Dependence of ISCO radius on $n$ for different values of $g_s$ and $q$ parameters. In the left panel, $q$ is fixed as $0.2$. In the right panel, $g_s$ is fixed as $0.2$ and $-0.2$.}
    \label{isco}
\end{figure}

Various factors influence the location of the ISCO, most notably the presence of scalar fields or non-gravitational interactions. When the interaction parameter $g_s$ is nonzero, the corresponding ISCO condition becomes substantially more intricate, reflecting the nonlinear coupling between the test particle and the background scalar field. In Fig.~\ref{isco}, we present the dependence of the ISCO radius on both the coupling parameter and the nonlinearity parameter, as obtained from our numerical analysis. Our numerical results show a clear asymmetry in the ISCO behavior depending on the sign of the coupling constant. In particular, the ISCO radius decreases for $g_s>0$, indicating that attractive scalar interactions enhance orbital stability at smaller radii, whereas for $g_s<0$ the ISCO shifts outward, signaling a destabilizing effect that pushes the marginally stable orbit to larger distances. These trends persist across a wide range of values of the nonlinearity parameter, demonstrating that scalar field effects can compete with and significantly modify the purely gravitational contribution. Such modifications of the ISCO structure have direct implications for accretion disk physics, including changes in the inner disk edge, radiative efficiency, and characteristic frequencies associated with quasiperiodic oscillations. Consequently, the ISCO provides a sensitive observational window into scalar field induced deviations from general relativity, making the study of scalar coupled compact objects a promising avenue in both theoretical investigations and astrophysical observations.

Radiative efficiency $\eta$ is defined by the relation:  
\begin{align}
\eta = 1 - \mathcal{E}_{\text{ISCO}} \tag{10}    
\end{align}
where $\mathcal{E}_{\text{ISCO}}$ denotes the specific energy of a particle orbiting in the innermost stable circular orbit (ISCO). Therefore, the efficiency $\eta$ depends on the properties of the underlying spacetime metric.

\begin{figure}
    \centering
    \includegraphics[width=0.45\linewidth]{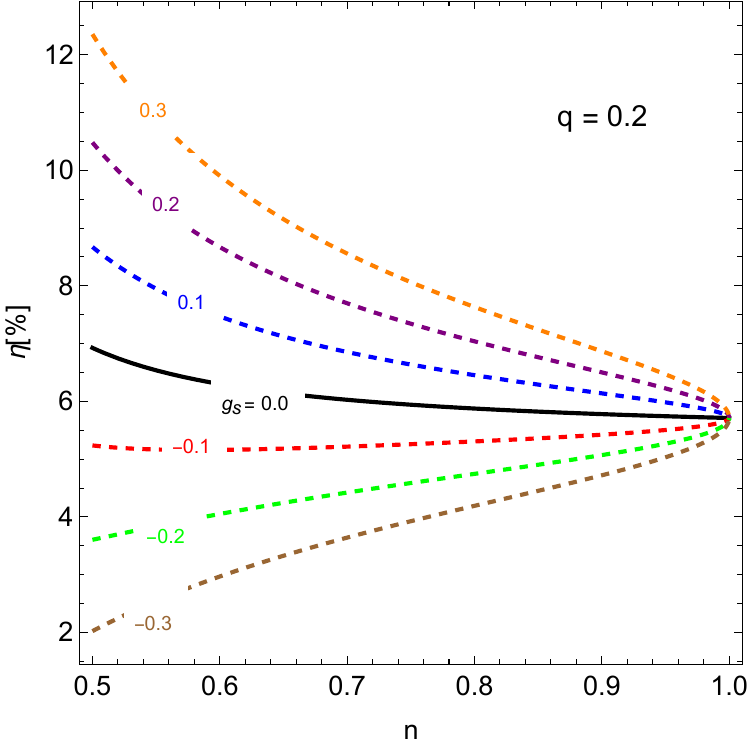}
    \includegraphics[width=0.45\linewidth]{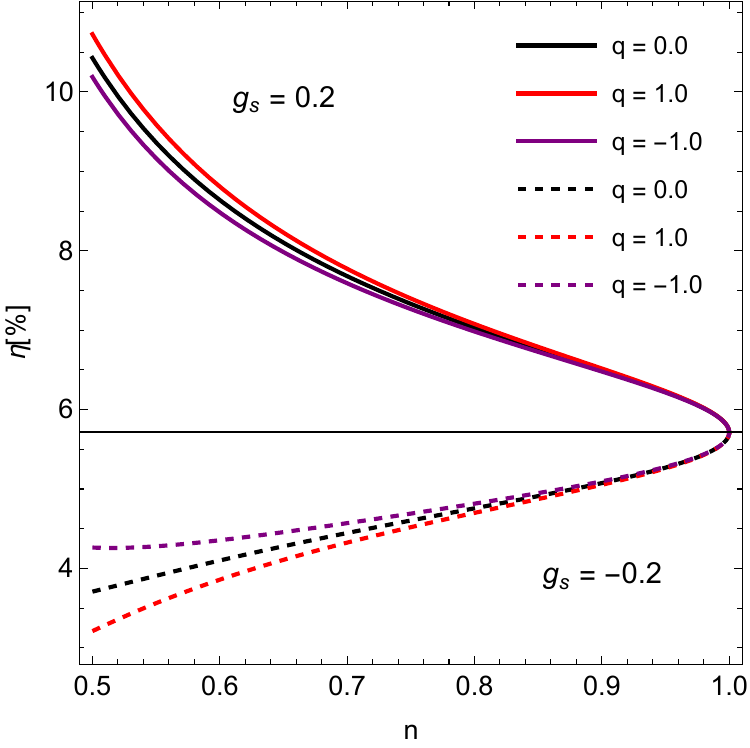}
    \caption{Dependence of the radiative efficiency $\eta$ on the parameter $n$ for different values of $g_s$ and $q$.}
    \label{figeta}
\end{figure}

Fig.~\ref{figeta} illustrates the dependence of efficiency $\eta$ on the parameter $n$ for various combinations of $g_s$ and $q$. Several key features can be observed. At $n=1$, all curves converge to the same efficiency value of approximately $6\%$, regardless of the parameters. For $n \neq 1$, positive values of $g_s$ enhance efficiency, while negative values reduce it. The parameter $q$ mainly affects the separation between the curves, leading to a splitting of the lines for different values of $q$. 

\subsection{Degeneracy Between Model Parameters and Kerr Spin}

In Kerr spacetime, the innermost stable circular orbit (ISCO) radius can be found analytically as a function of the black hole spin parameter $a$ \cite{bardeen1972}:

\begin{align}
r_{\mathrm{ISCO}} = 3 + Z_2 \pm \sqrt{(3-Z_1)(3+Z_1+2Z_2)},
\end{align}
where the upper sign corresponds to prograde orbits and the lower sign to retrograde orbits, and the auxiliary functions are given by

\begin{align}
Z_1 &= 1 + \left(\sqrt[3]{1+a} + \sqrt[3]{1-a}\right) \sqrt[3]{1-a^2},\nonumber\\
Z_2 &= \sqrt{3a^2 + Z_1^2}.
\end{align}
For a Schwarzschild black hole ($a=0$), this reduces to the familiar $r_{\mathrm{ISCO}} = 6M$, while for a maximally rotating Kerr black hole ($a=1$), the prograde ISCO approaches $r_{\mathrm{ISCO}} = M$.

A key question in testing alternative gravity theories is whether the parameters of a given model can produce observational signatures that mimic those of the Kerr spin parameter. In the proposed scalar-tensor framework, we investigate how the parameters $q$, $n$, and $g_s$ can collectively reproduce the ISCO shifts typically attributed to black hole spin in general relativity. This degeneracy is particularly important for interpreting QPO observations, as different combinations of these parameters can yield identical epicyclic frequency profiles.

\begin{figure}
\centering
\includegraphics[width=0.32\linewidth]{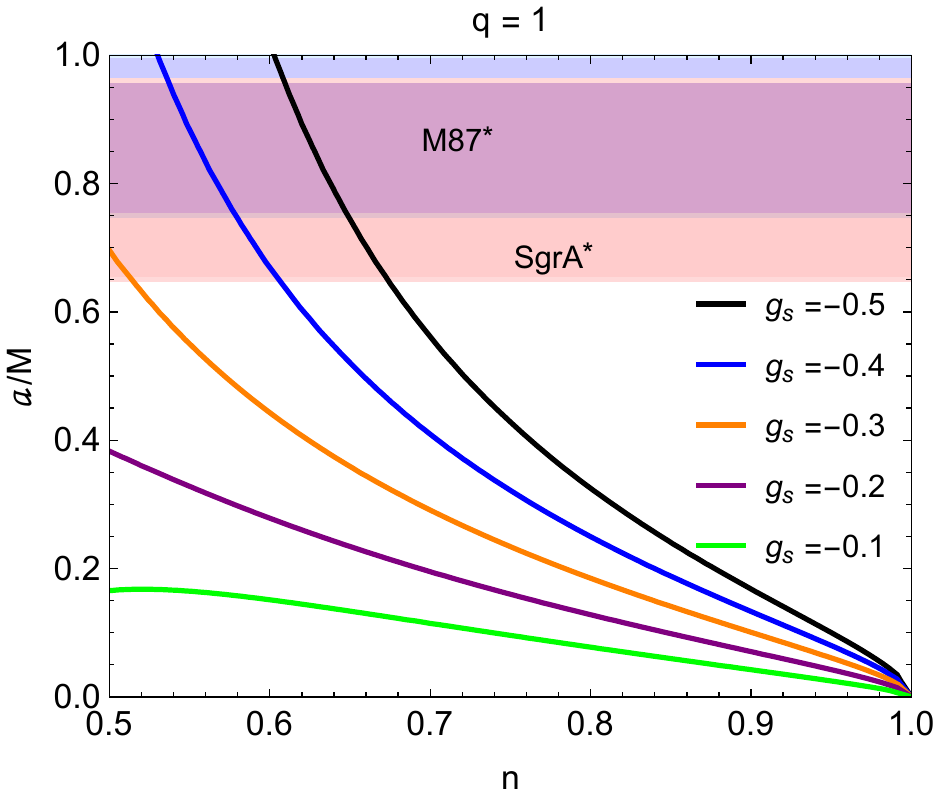}
\includegraphics[width=0.32\linewidth]{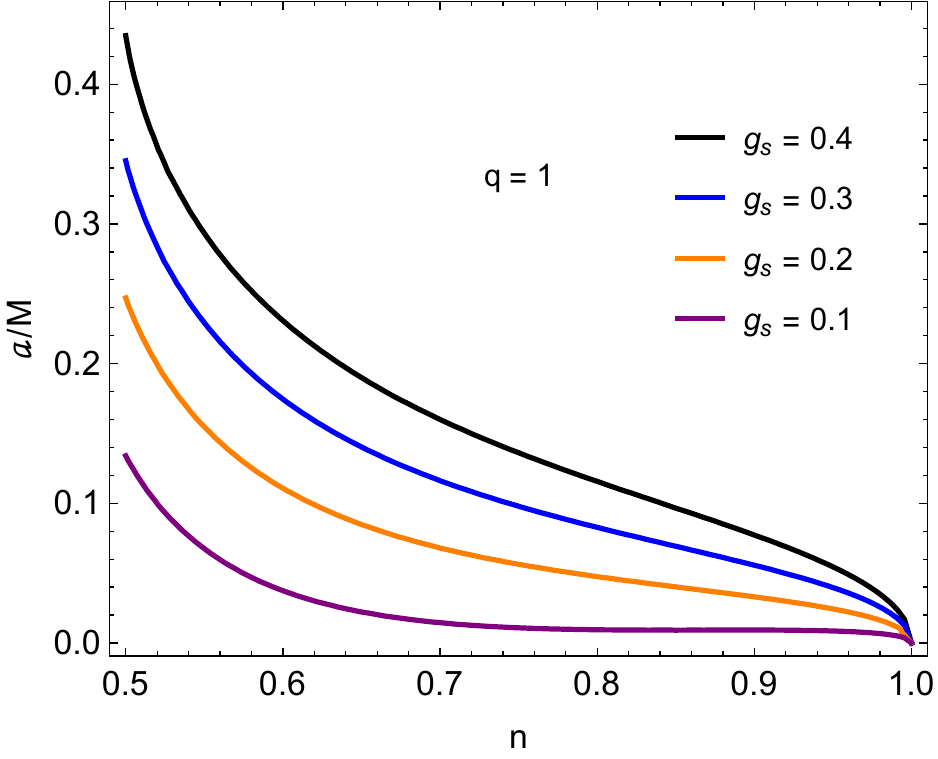}
\includegraphics[width=0.32\linewidth]{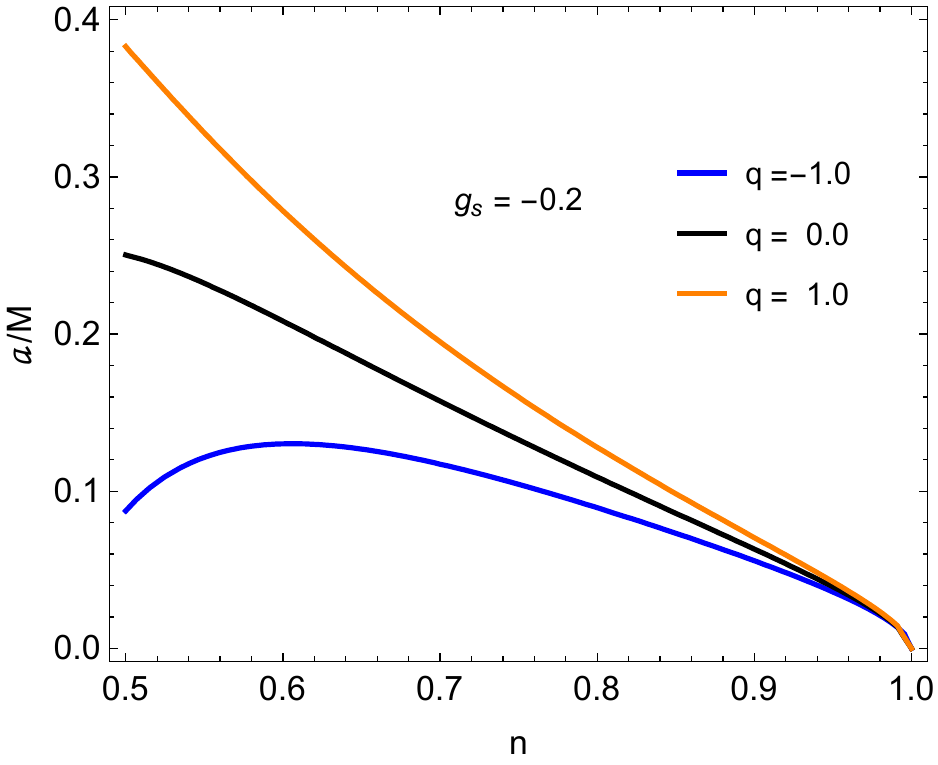}
\caption{Degeneracy between the scalar-tensor model parameters and the Kerr spin parameter $a$. Each panel shows the ISCO radius as a function of the Kerr spin $a$ (solid curve), along with the ISCO values obtained for different combinations of $q$, $n$, and $g_s$. The intersection points indicate parameter values that produce the same ISCO location as a given Kerr spin.}
\label{mimic}
\end{figure}

Fig.~\ref{mimic} illustrates how the parameters of the considered model can mimic the effect of the Kerr spin parameter. The solid curve represents the dependence of the ISCO radius on the spin parameter $(a)$ in Kerr spacetime, while the horizontal lines (or discrete points) correspond to the ISCO radii obtained for different combinations of the parameters $(q)$, $(n)$, and ($g_s$) in our scalar–tensor solution.

The intersection points between the Kerr curve and these lines indicate parameter degeneracies. In particular, small values of the spacetime parameter $(n)$ can produce shifts in the ISCO radius comparable to those generated by a rapidly rotating Kerr black hole, even when the intrinsic spin is absent. This behavior demonstrates that careful multi-frequency QPO modeling is required in order to break such degeneracies and reliably determine the underlying spacetime parameters.

Moreover, for certain combinations of the model parameters, the effective ISCO radius corresponds to Kerr spin values comparable to those inferred for the supermassive black holes at the centers of the Milky Way and $\rm{M87}$, namely $(\mathrm{Sgr~A^*})$ with $(a/M \sim 0.65 - 0.9)$ and $(\mathrm{M87^*})$ with $(a/M \sim 0.75 - 1)$ \cite{EHTM87_2019,EHTSgrA_2022}.

\section{Oscillatory Motion}\label{sec.6}

In this section, we will concentrate on the oscillatory motion of a massive particle near a stable circular orbit within the JNW spacetime, evaluating the influence of a scalar field. In this scenario, the four-velocity of the particle can be expressed as: 

$$u^\mu = \dot{t}(1, 0, 0, \Omega)\ ,$$ 
where $\Omega = \frac{d\phi}{dt}$ represents the angular velocity of the particle. The radial equation simplifies to the following:

\begin{align}\label{W}
\Omega^2 = -\frac{(1 + g_s \varphi) \partial_r g_{tt} - 2 g_s g_{tt} \varphi'(r)}{(1 + g_s \varphi) \partial_r g_{\phi\phi} - 2 g_s g_{\phi\phi} \varphi'(r)}.
\end{align}

It is important to note that when the scalar field is absent, the explicit formula for the angular velocity of a massive particle in JNW spacetime simplifies to 

$$\Omega = \Omega_K f^{n - 1/2} \left[1 - M(1 + n)/nr\right]^{-1}.$$ 

As a result, the expressions for the specific energy and angular momentum of the massive particle are given by:

\begin{align}
\mathcal{E} = -(1 + g_s \varphi) g_{tt} \dot{t} = -\frac{(1 + g_s \varphi) g_{tt}}{\sqrt{-g_{tt} - \Omega^2 g_{\phi\phi}}}, \\
\\
\mathcal{L} = (1 + g_s \varphi) g_{\phi\phi} \Omega \dot{t} = \frac{(1 + g_s \varphi) g_{\phi\phi} \Omega}{\sqrt{-g_{tt} - \Omega^2 g_{\phi\phi}}}.
\end{align}

By applying the normalization condition for the four-velocity of the massive particle, we can derive:

\begin{align}\label{2D}
g_{rr} \dot{r}^2 + g_{\theta\theta} \dot{\theta}^2 + V(r, \theta) = 0,
\end{align}
where $V(r, \theta)$ is the effective potential detailed in equation \eqref{eom}. From equation \eqref{2D}, we can derive the equation for a 2D harmonic oscillator by expanding it up to the leading order:

\begin{align}\nonumber
&g_{rr}(\dot{r}_0 + \dot{\delta r})^2 + g_{\theta\theta}(\dot{\theta}_0 + \dot{\delta \theta})^2 + V(r_0, \theta_0) \\ \nonumber
&+ \partial_r V(r_0, \theta_0) \delta r + \partial_\theta V(r_0, \theta_0) \delta \theta + \partial_r \partial_\theta V(r_0, \theta_0) \delta r \delta \theta \\
&+ \frac{1}{2} \partial_r^2 V(r_0, \theta_0) \delta r^2 + \frac{1}{2} \partial_\theta^2 V(r_0, \theta_0) \delta \theta^2 + ... = 0.
\end{align}

Consequently, the equations governing the oscillatory motion for the radial and vertical displacements $\delta r$ and $\delta \theta$ from the stationary point are given by:

\begin{align}\nonumber
&g_{rr} \left( \frac{d^2}{dt^2} + \Omega_r^2 \right) \delta r + g_{\theta\theta} \left( \frac{d^2}{dt^2} + \Omega_\theta^2 \right) \delta \theta = 0,
\end{align}
where
\begin{align}
&\Omega_r^2 = \frac{1}{2 g_{rr} \dot{t}^2} \partial_r^2 V = \frac{-g_{tt} - \Omega_{\phi}^2 g_{\phi\phi}}{2 g_{rr}} \partial_r^2 V, \\
\\
\label{Wq}
&\Omega_\theta^2 = \frac{1}{2 g_{rr} \dot{t}^2} \partial_\theta^2 V = \frac{-g_{tt} - \Omega_{\phi}^2 g_{\phi\phi}}{2 g_{\theta\theta}} \partial_\theta^2 V.
\end{align}

The analysis performed indicates that the vertical frequency in equation \eqref{Wq} matches the orbital angular velocity in equation \eqref{W} (i.e., $\Omega_\theta = \Omega$). Throughout the analysis, we express all fundamental frequencies in Hz, thereby reintroducing fundamental constants into all frequencies as follows:
$\nu_i = \frac{1}{2\pi} \frac{c^3}{G M} \Omega_i\ ,$
where $c = 3 \times 10^{10} \, \text{cm s}^{-1}$ is the speed of light in a vacuum and $G = 6.67 \times 10^{-8} \, \text{cm}^3 \text{g}^{-1} \text{s}^{-2}$ is the Newtonian gravitational constant.

\begin{figure}
    \centering
    \includegraphics[width=0.32\linewidth]{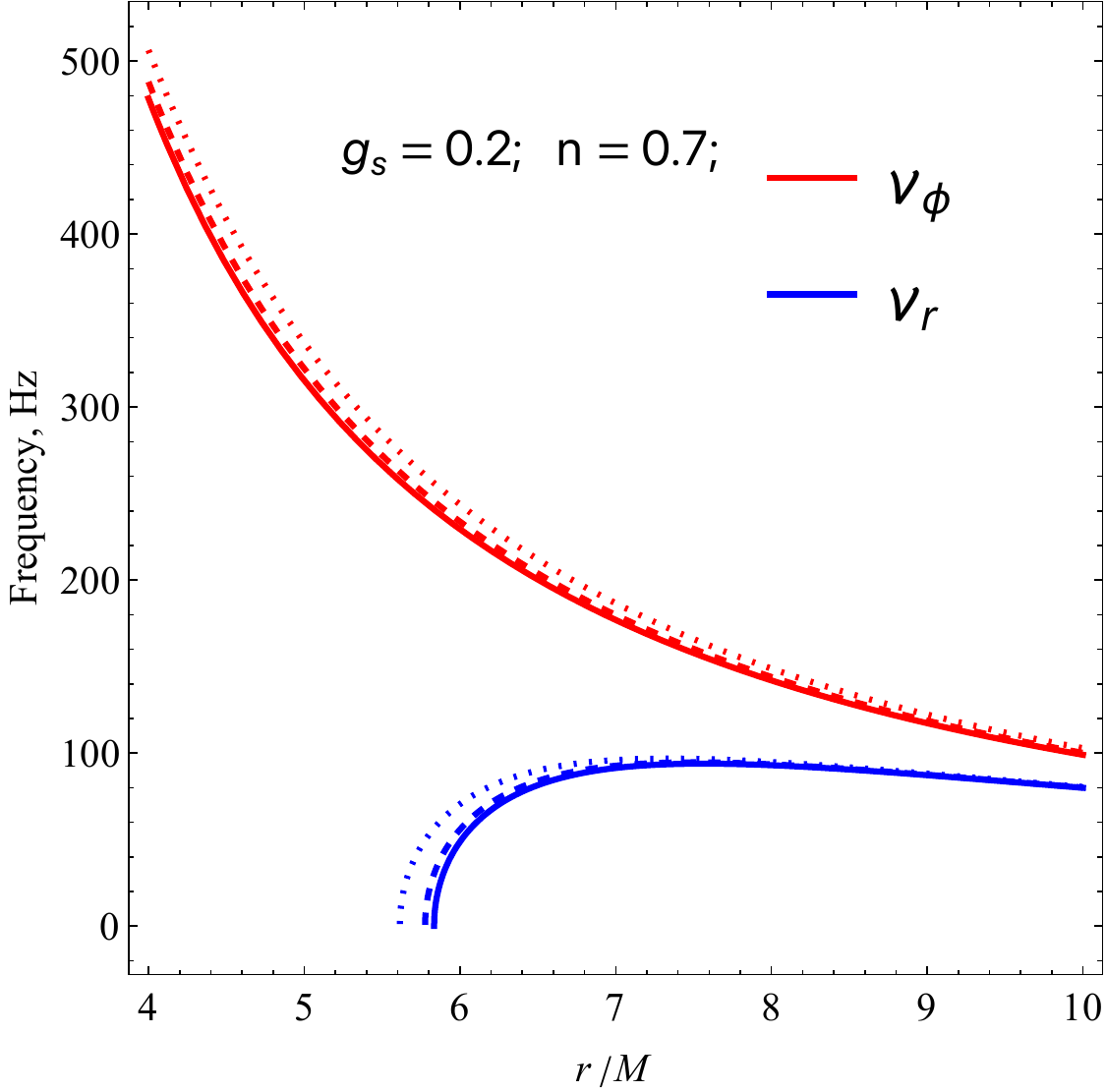}
    \includegraphics[width=0.32\linewidth]{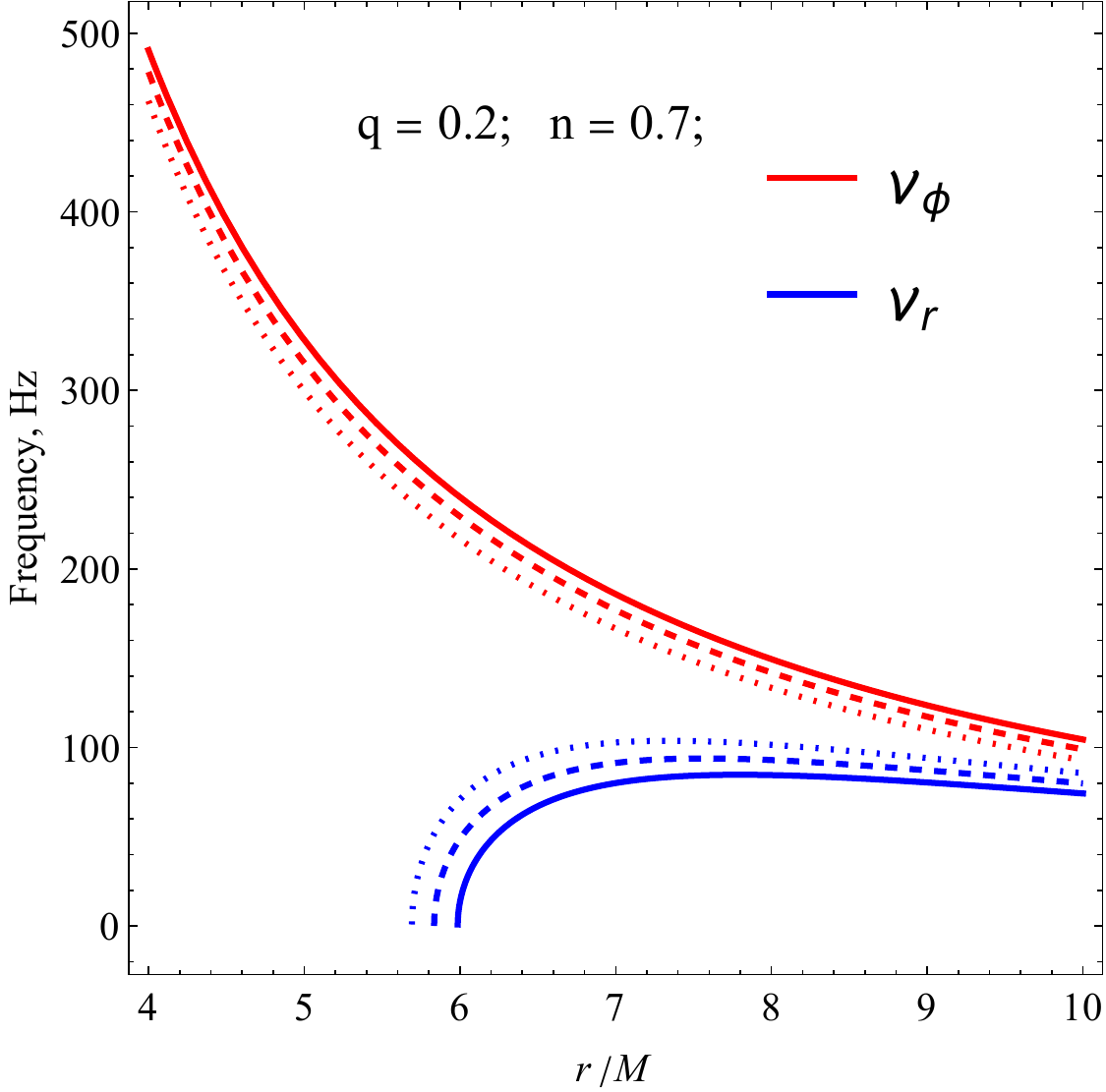}
    \includegraphics[width=0.32\linewidth]{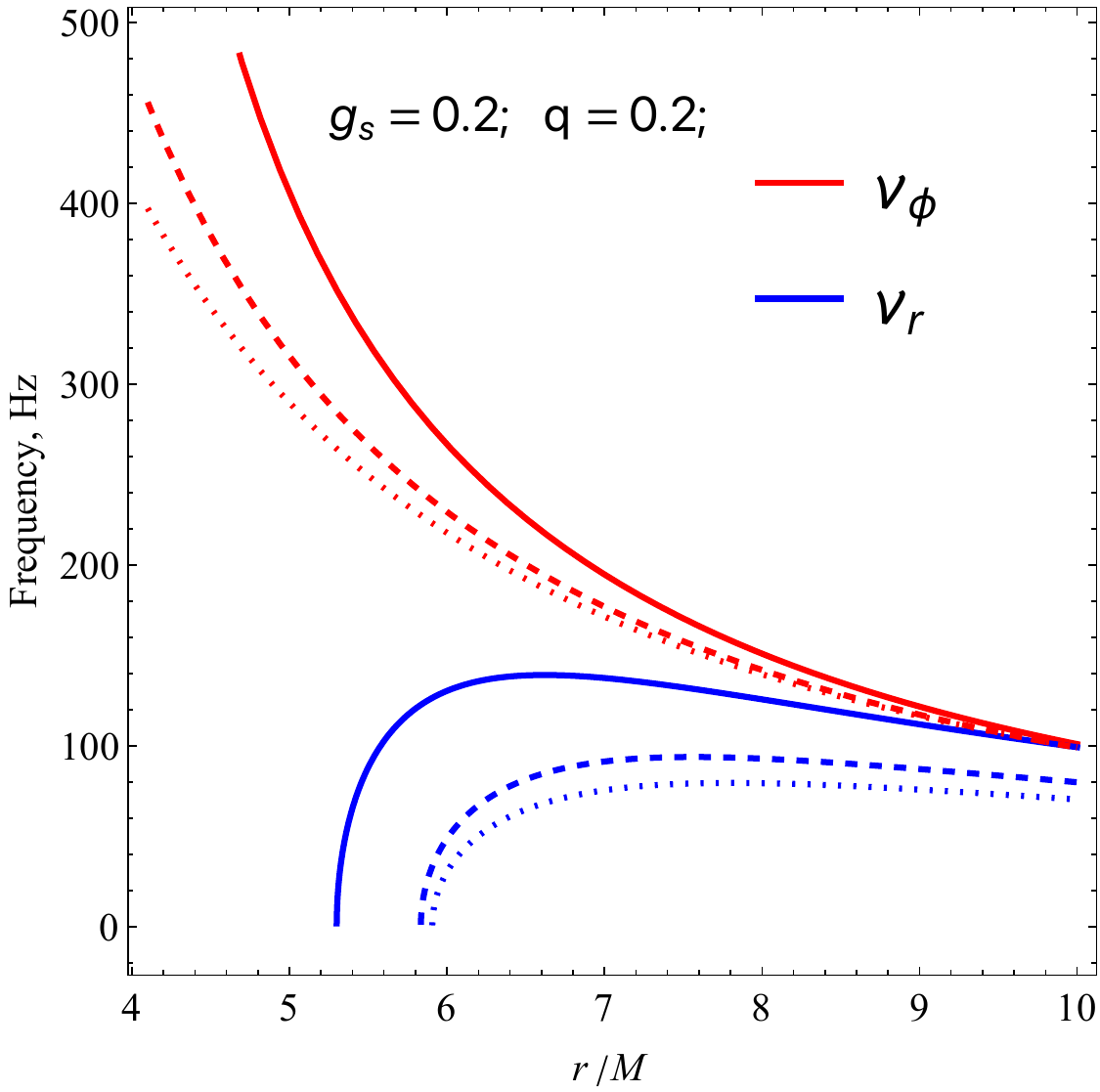}
    \caption{Radial dependence of the fundamental frequencies $\nu_r$ and $\nu_\phi$. Left panel: Fixed $g_s = 0.2$ and $n = 0.7$, with $q = 0.2$ (solid), $1.0$ (dashed), and $3.0$ (dotted). Middle panel: Fixed $q = 0.2$ and $n = 0.7$, with $g_s = 0.1$ (solid), $0.2$ (dashed), and $0.3$ (dotted). Right panel: Fixed $g_s = 0.2$ and $q = 0.2$, with $n = 0.5$ (solid), $0.7$ (dashed), and $0.9$ (dotted).}
    \label{figfr}
\end{figure}

Fig.~\ref{figfr} shows the radial dependence of the epicyclic frequencies $\nu_r$ and $\nu_\phi$. In a spherically symmetric spacetime, we have $\nu_\theta = \nu_\phi$, so the latter is not shown separately. From the left panel, we observe that increasing the parameter $q$ causes a slight increase in the frequency $\nu_r$ and systematically shifts its peak toward smaller radii. The middle panel illustrates the effect of $g_s$ for fixed $n$ and $q$. Here, increasing $g_s$ leads to an increase in $\nu_r$ but a decrease in $\nu_\phi$. The right panel demonstrates the influence of the spacetime parameter $n$, which has a relatively larger effect compared to the other parameters. Increasing $n$ causes a decrease in both $\nu_r$ and $\nu_\phi$.

\subsection{The epicyclic resonance (ER) model}\label{sec5.1}

The epicyclic resonance (ER) model provides an explanation for the quasi-periodic oscillations (QPOs) observed in accretion disks around compact objects, such as those described by the Schwarzschild and Kerr metrics. In this framework, the QPOs arise from a non-linear resonance between the epicyclic modes of oscillating fluid elements in the disk \cite{Kluniak2002ParametricER, Banerjee2022TestingBH}.

Specifically, the upper frequency $\nu_U$ is associated with the azimuthal (orbital) frequency $\nu_\phi$, while the lower frequency $\nu_L$ corresponds to the radial epicyclic frequency $\nu_r$:

\begin{align}
    \nu_U = \nu_{\phi}; \quad \nu_L = \nu_r.
\end{align}

This model helps in interpreting the twin-peak QPOs by linking them to the fundamental frequencies of geodesic motion in the gravitational potential of the black hole.

\begin{figure}
    \centering
    \includegraphics[width=0.32\linewidth]{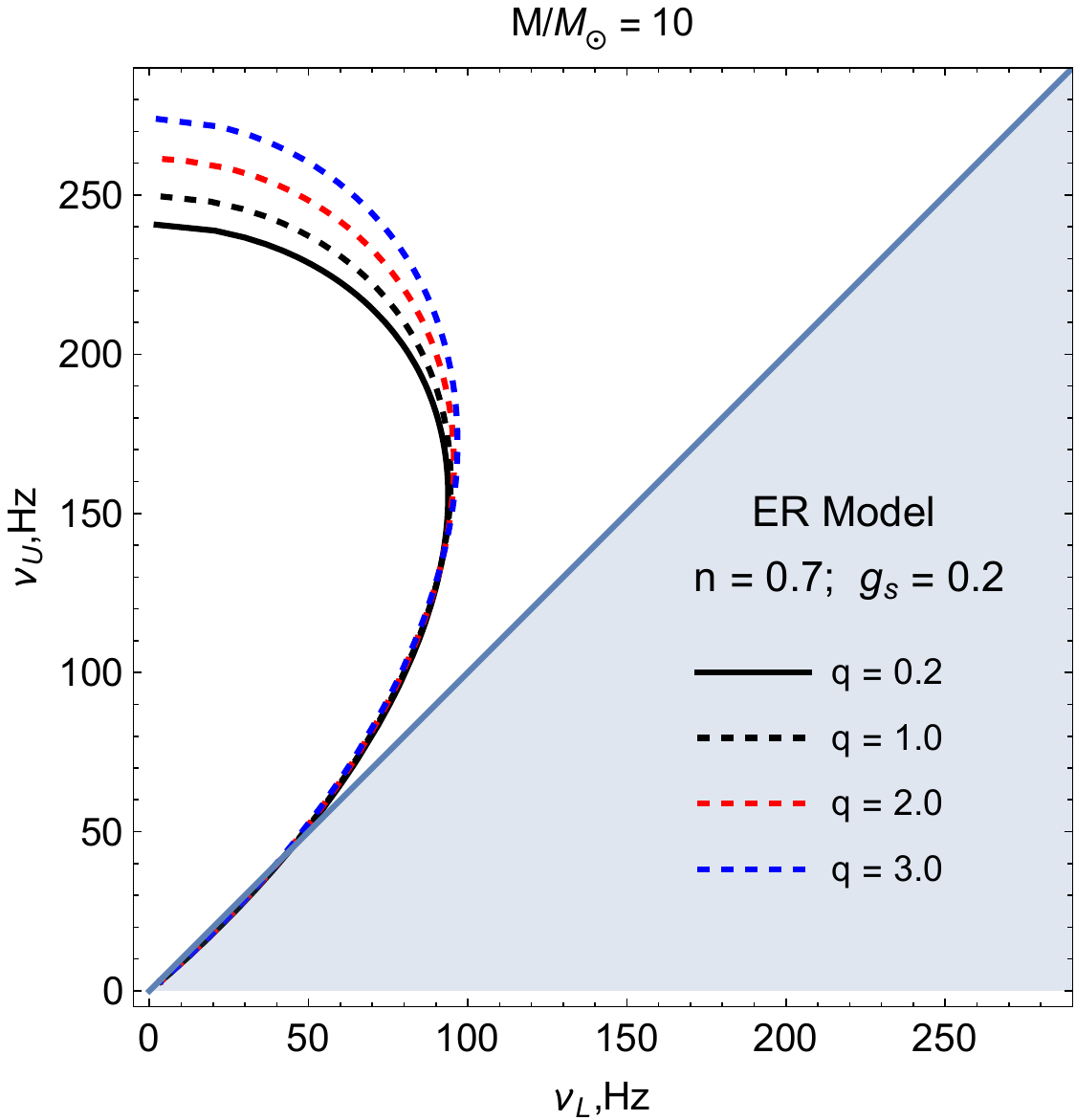}
    \includegraphics[width=0.32\linewidth]{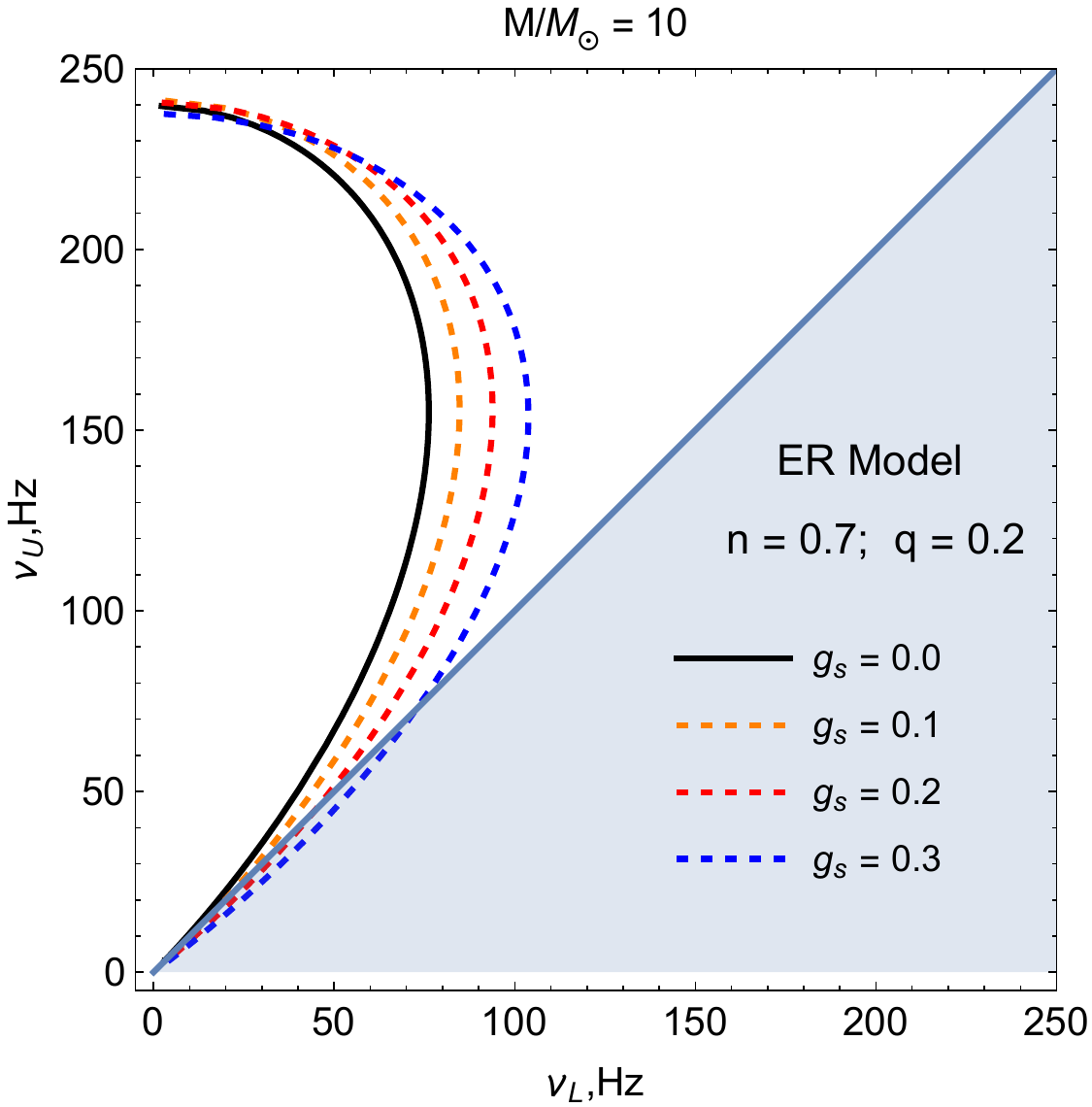}
    \includegraphics[width=0.32\linewidth]{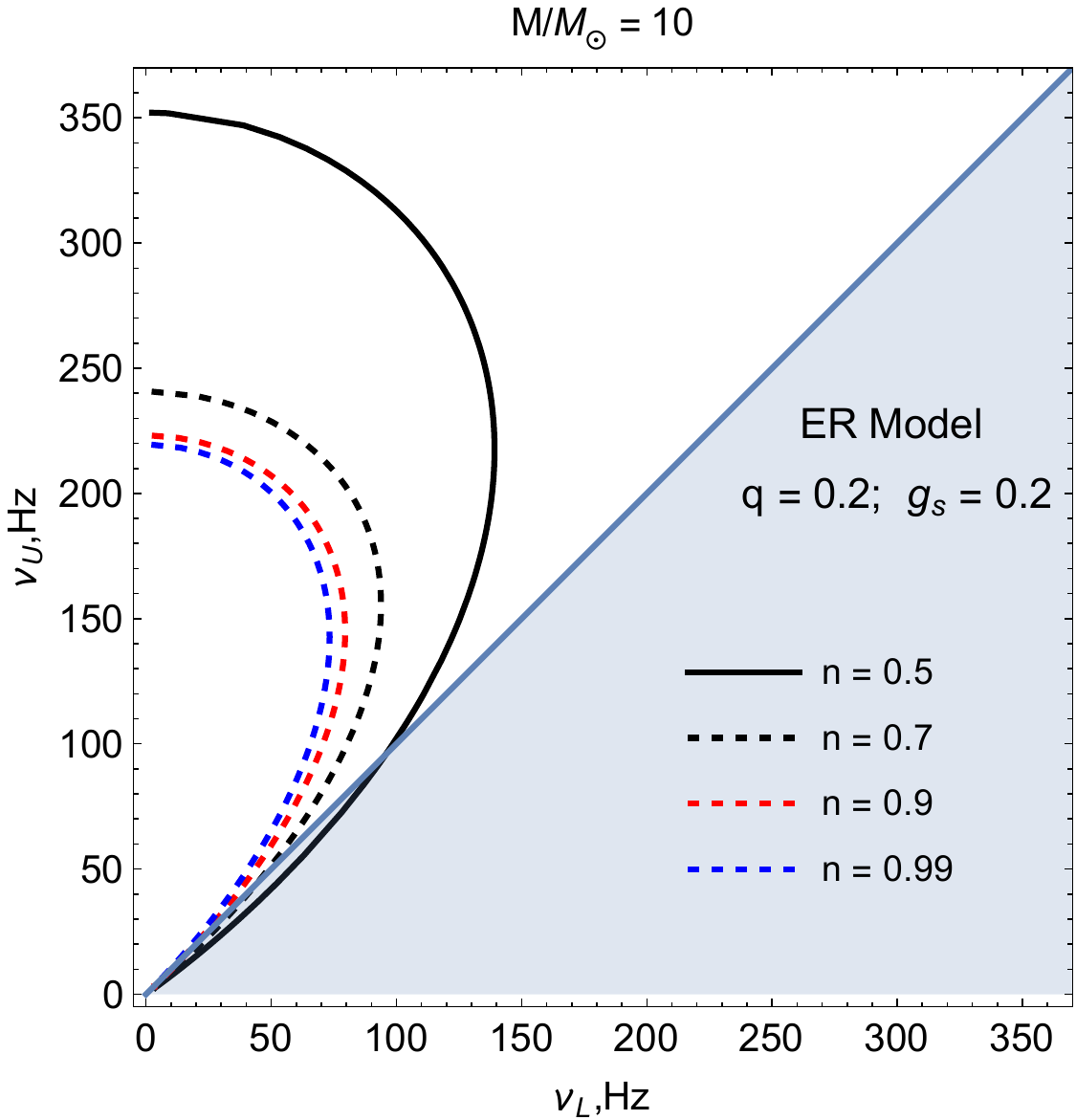}    
    \caption{The relationship between the high-frequency ($\nu_U$) and low-frequency ($\nu_L$) peaks of twin-peak QPOs in the ER model, assuming a black hole mass of $M = 10 M_{\odot}$.}
    \label{fig:ul}
\end{figure}

Fig.~\ref{fig:ul} shows the parametric dependence of the upper and lower frequencies within the ER model. The three panels illustrate the distinct effects of each parameter. In the left panel, for fixed values of $n$ and $g_s$, increasing $q$ leads to a noticeable increase in the upper frequency. The middle panel demonstrates the effect of $g_s$, which is different from that of $q$; increasing $g_s$ primarily raises the lower frequency. The right panel shows the influence of $n$, which significantly affects both the upper and lower frequencies, with an overall impact that is comparatively larger than that of the other parameters.

\section{Constraint}\label{sec.7}

\begin{figure}
    \centering
    \includegraphics[width=0.5\linewidth]{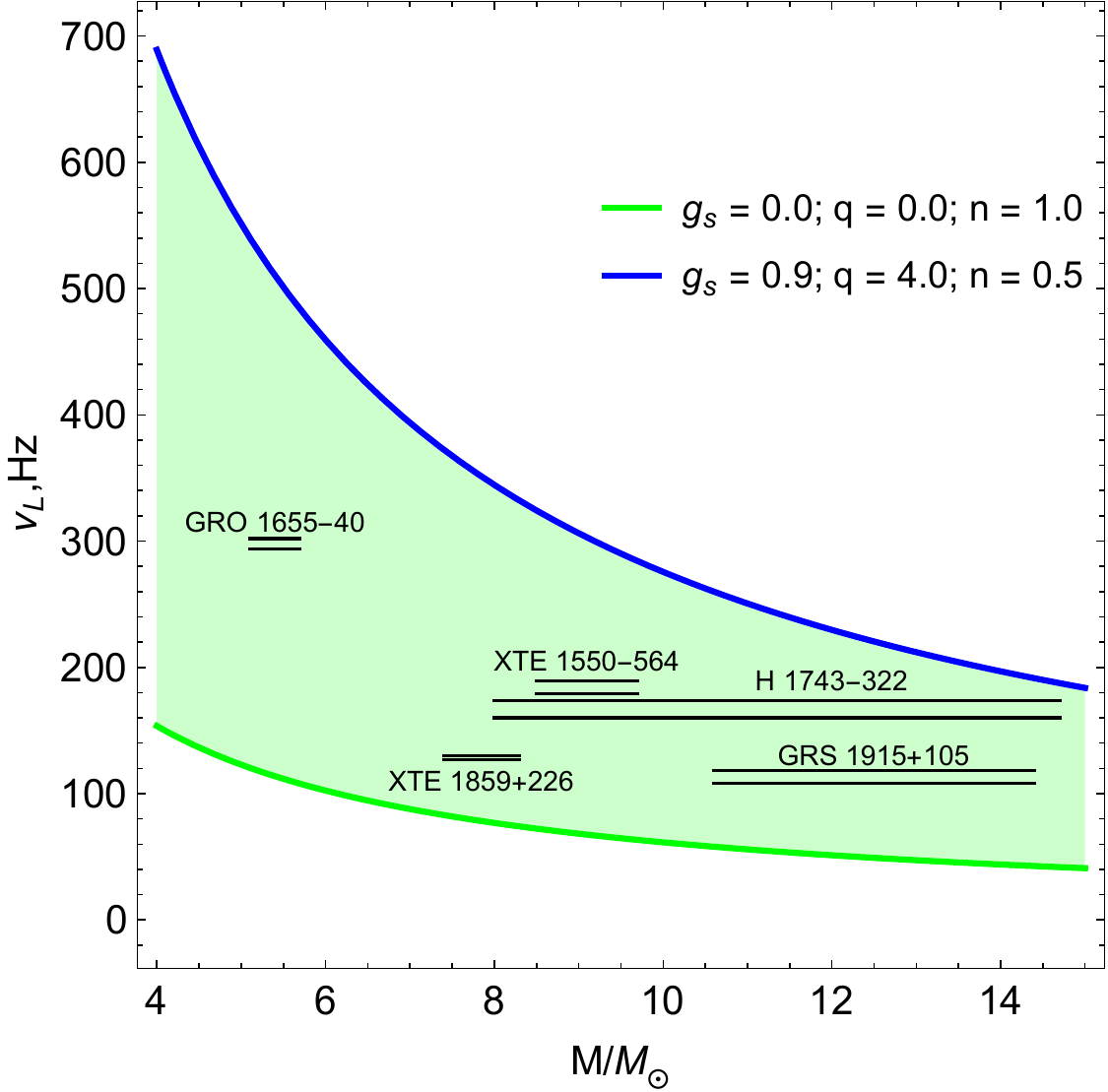}
    \caption{Dependence of the lower frequencies on mass (in solar mass units). The blue curve corresponds to the parameters ( $g_s = 0$ ), ( $q = 0$ ), and ( $n = 1.0$ ), whereas the green curve represents ( $g_s = 0.9$ ), ( $q = 4.0$ ), and ( $n = 0.5$ ). The shaded area marks the region of observed data, indicating that the model is consistent with these observations.}
    \label{data}
\end{figure}

In this section, we examine five X-ray binary systems to impose constraints on the parameters of our Janis-Newman-Winicour (JNW) model, leveraging Quasi-Periodic Oscillation (QPO) data derived from these systems. The specific sources under investigation include XTE J1550-564\cite{Remillard:2002cy}, GRO J1655-40\cite{Motta:2013wga}, XTE 1859+226\cite{Motta:2022rku}, H 1743-322\cite{Ingram:2014ara}, and GRS 1915+105\cite{Remillard:2006fc}. Their measured frequencies are detailed in Table \ref{table1}. Employing the upper and lower frequency values, we demonstrate in Fig.~\ref{data} the extent to which we can characterize the observational data from these sources. That said, XTE 1859+226,  H 1743-322 and GRO J1655-40 display nodal frequencies, signifying that they involve rotation and thus cannot be adequately modeled using our static black hole framework. Consequently, we will report the best-fit parameter values for the other three sources—namely, GRS 1915+105 and XTE J1550-564—within the explored parameter space, as determined through Markov Chain Monte Carlo (MCMC) simulations.

\begin{table}[h!]
\centering
\caption{The mass, orbital frequencies, periastron precision frequencies and nodal precision frequencies of QPOs from the X-ray binaries selected for analysis.} 
\begin{tabular}{||c c c c c c||} 
 \hline
 &GRO J1655-40 & XTE J1550-564 & XTE J1859+226 & GRS 1915+105 & H1743-322 \\ [0.6ex] 
 \hline\hline
 $M(M_{\odot})$ & 5.4$\pm$0.3 & 9.1 $\pm$ 0.61 & 7.85$\pm$0.46  & $12.4^{+2.0}_{-1.8}$ & $\gtrsim$ 9.29 \\ 
 $\nu_{U}$(Hz) & 441$\pm$ 2 & 276 $\pm$ 3 & $227.5^{+2.1}_{-2.4}$ & 168 $\pm$ 3 & 240 $\pm$ 3\\
 $\nu_{L}$(Hz) & 298 $\pm$ 4 & 184 $\pm$ 5 & $128.6^{+1.6}_{-1.8}$ & 113 $\pm$ 5 & $165^{+9}_{-5}$ \\
 $\nu_{nod}$(Hz) & 17.3 $\pm$ 0.1 & - & 3.65 $\pm$ 0.01 & - & 9.44 $\pm$ 0.02 \\[1ex] 
 \hline
\end{tabular}
\label{table1}
\end{table}

For the MCMC analysis, we employed the Python package emcee \cite{2013PASP125306F} to constrain the JNW parameters using the epicyclic resonance (ER) framework.

The posterior distribution is formulated as described in \cite{Liu:2023vfh}:
\begin{align}\label{Dist}
\mathcal{P}(\Theta|D,M) = \frac{P(D|\Theta,M)\pi(\Theta|M)}{P(D|M)} \quad ,
\end{align}
where:
- $\pi(\Theta)$ denotes the prior distribution,
- $P(D|\Theta, M)$ represents the likelihood function.

The priors follow a Gaussian form confined to defined limits, specifically:
\begin{align}\label{pitheta}
\pi(\Theta_i) \sim \exp\left( -\frac{1}{2\sigma_i^2} (\Theta_i - \Theta_0)^2 \right) ,
\end{align}
with $\Theta_{\text{low},i} < \Theta_i < \Theta_{\text{high},i}$ for the parameters $\Theta_i = [M, g_s, n, q, r/M]$, and $\sigma_i$ indicating the respective standard deviations. 

Drawing from the upper and lower frequency observations presented in Section \ref{sec5.1}, our MCMC procedure integrates two distinct datasets. Central to this is the likelihood function $\mathcal{L}$, which is given by:
\begin{align}\label{LogL}
\log \mathcal{L} = \log \mathcal{L}_{U} + \log \mathcal{L}_{L} ,
\end{align}
The logarithmic likelihood is thus the aggregate of the logs for the upper and lower frequency components. The term $\log \mathcal{L}_{U}$ corresponds to the likelihood of the upper-frequency observations, expressed as:
\begin{align}\label{LogL_up}
\log \mathcal{L}_{U} = -\frac{1}{2} \sum_{i} \left( \frac{(\nu^{i}_{\text{U, obs}} - \nu^{i}_{\text{U, th}})^2}{(\sigma^{i}_{\text{U, obs}})^2} \right) ,
\end{align}
Similarly, $\log \mathcal{L}_{L}$ pertains to the lower-frequency observations, formulated as:
\begin{align}\label{LogL_low}
\log \mathcal{L}_{L} = -\frac{1}{2} \sum_{i} \left( \frac{(\nu^{i}_{\text{L, obs}} - \nu^{i}_{\text{L, th}})^2}{(\sigma^{i}_{\text{L, obs}})^2} \right) ,
\end{align}
Here, the index $i$ spans from 1 to the total count of available upper and/or lower frequency measurements. The terms $\nu^i_{\text{U, obs}}$ and $\nu^i_{\text{L, obs}}$ signify the empirical upper and lower frequencies, labeled as $\nu_{U}$ and $\nu_{L}$, while $\nu^i_{\text{U, th}}$ and $\nu^i_{\text{L, th}}$ are their theoretical counterparts. Moreover, $\sigma^i_{U}$ and $\sigma^i_{L}$ capture the associated statistical errors.

\begin{figure}
    \centering
    \includegraphics[width=0.7\linewidth]{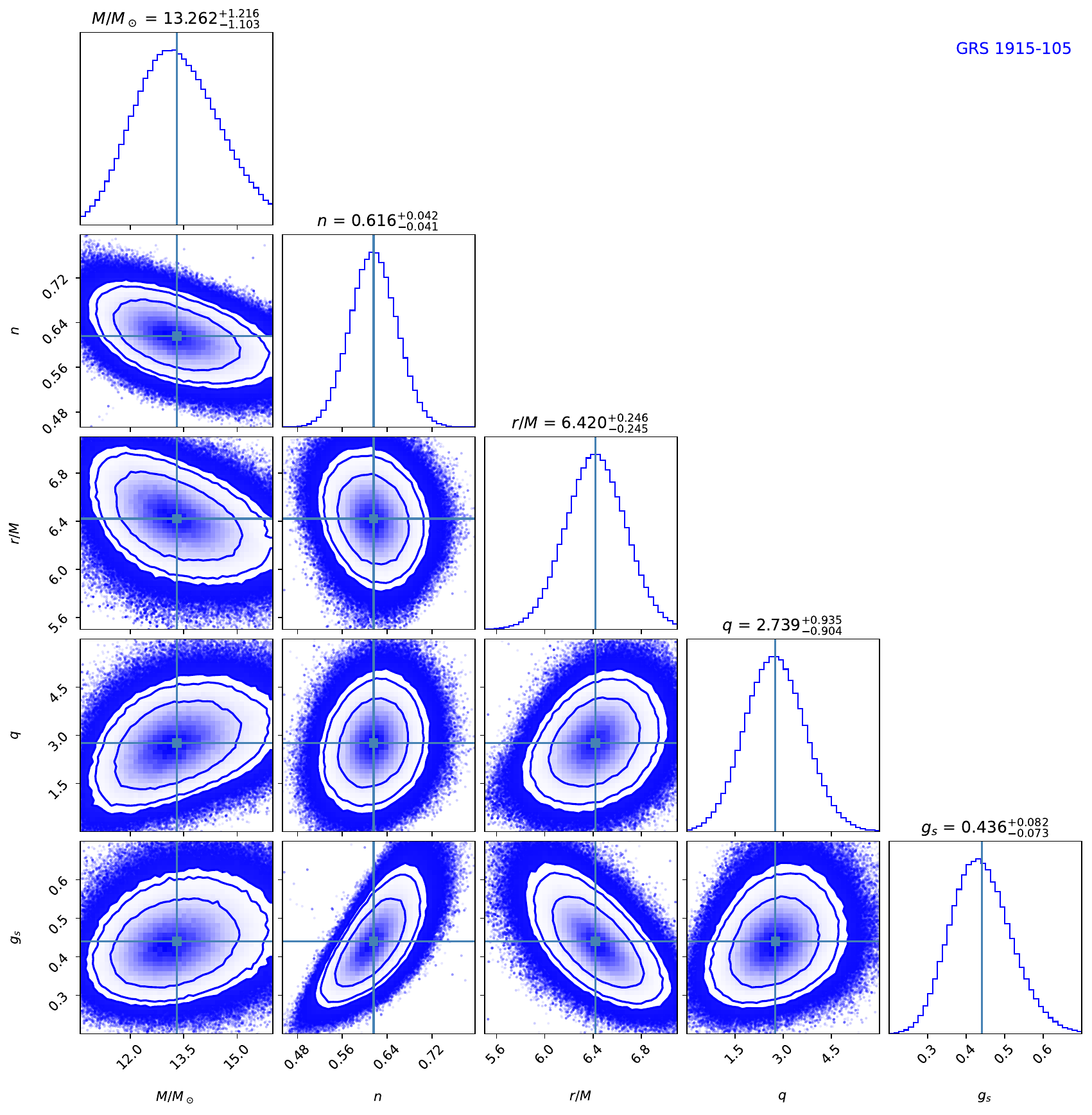}
    \centering\includegraphics[width=0.7\linewidth]{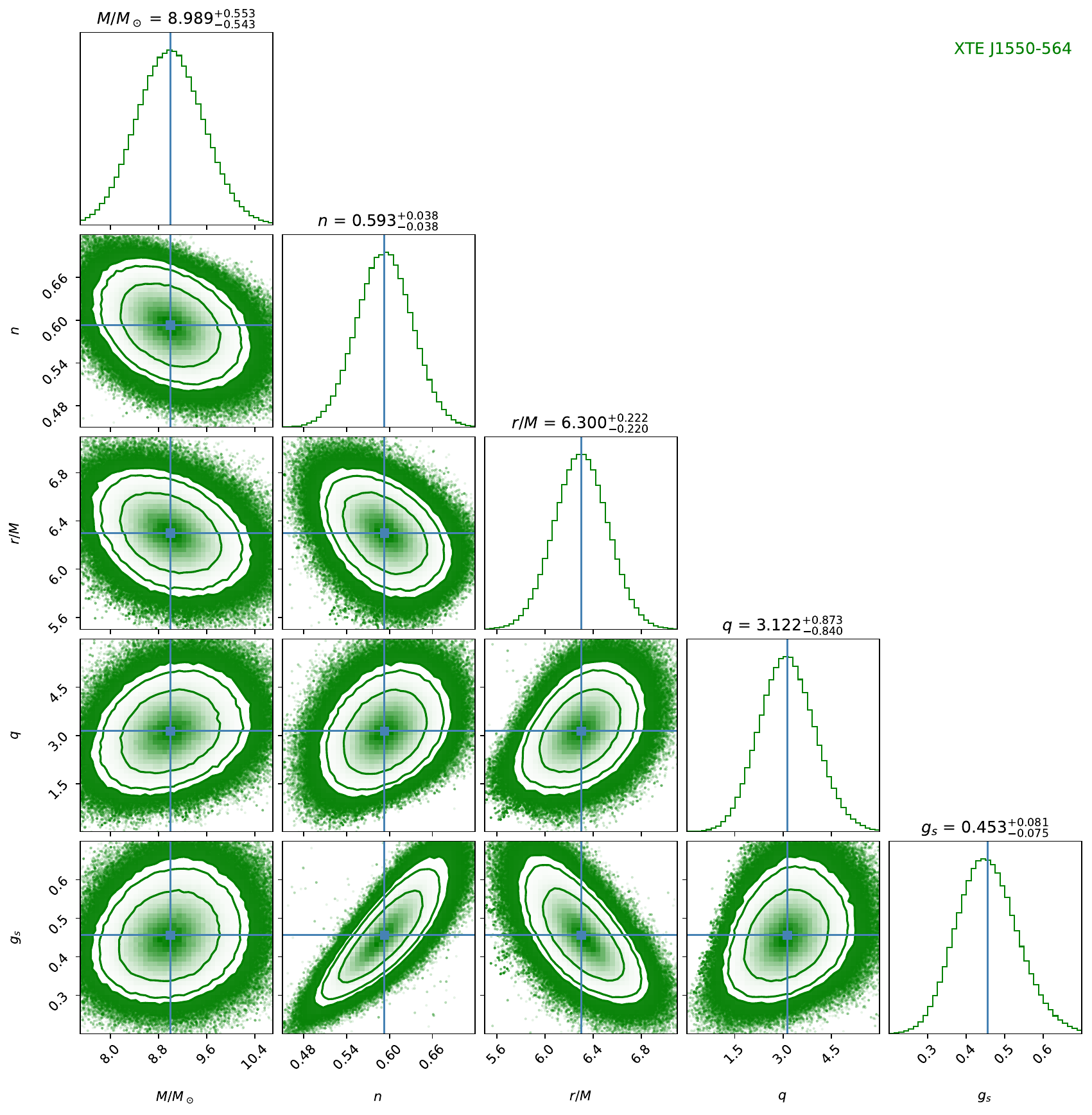}
    \caption{Bounds on the mass in the Janis-Newman-Winicour (JNW) framework, as well as the $g_s$, $q$, and $n$ parameters, for the microquasars GRS 1915+105 (upper panel) and XTE J1550-564 (lower panel), derived via Markov Chain Monte Carlo (MCMC) techniques.}
    \label{figmcmc}
\end{figure}

Through MCMC evaluations at the $1\sigma$, $2\sigma$ and $3\sigma$ confidence intervals, as illustrated in the corner plots of Fig.~\ref{figmcmc} for the three specified sources, we depict the optimal parameter estimates ($M, g_s, n, q, r$) in 1$\sigma$ aligned with the JNW metric. These optimal values are summarized in Table \ref{table2}.

\begin{table}[]
    \centering
    \begin{tabular}{||c c c||} 
 \hline
 & XTE J1550-564 & GRS 1915+105\\ [0.6ex] 
 \hline\hline
 $M(M_{\odot})$ & $8.989^{+0.553}_{-0.543}$ & $13.262^{+1.216}_{-1.103}$\\
 $n$ &$0.593\pm0.038$ &$0.616^{+0.042}_{-0.041}$\\
 $r/M$ & $6.300^{+0.222}_{-0.220}$&$6.420^{+0.246}_{-0.245}$ \\
 $q$ &$3.122^{+0.873}_{-0.840}$ & $2.739^{+0.935}_{-0.904}$\\
 $g_s$& $0.453^{+0.081}_{-0.075}$& $0.436^{+0.082}_{-0.073}$ \\[1ex] 
 \hline
\end{tabular}
    \caption{The optimal parameter estimates aligned with the Janis-Newman-Winicour (JNW) framework, inferred from Quasi-Periodic Oscillations (QPOs) in the selected X-ray binary systems.}
    \label{table2}
\end{table}

\section{Conclusions\label{Sec:Conclusions}}

In this work, we conducted a comprehensive study of the properties of a solution of a modified scalar-tensor gravity model and its astrophysical implications for black hole observations. Our key findings and contributions are summarized below.

\begin{itemize}
    \item We have derived an exact, static, spherically symmetric solution within the framework of Freud-Nambu scalar-tensor gravity. This solution represents a novel generalization of the well-known Janis-Newman-Winnicour (JNW) naked singularity spacetime, distinguished by the introduction of a new parameter $q$. A crucial feature of the derived solution is that the modification appears exclusively in the scalar field profile; the line element remains identical to that of the JNW metric. The JNW solution is consistently recovered in the limit $q = 0$, confirming the result obtained as a natural extension of this important class of spacetimes. This demonstrates the possibility of obtaining parametrized modifications of scalar-tensor gravity solutions, which could be extended to other theoretical frameworks.

    \item We conducted a detailed analysis of particle dynamics in the obtained spacetime. In addition to the standard geodesic motion, we introduced a direct linear coupling between the test particle and the scalar field, characterized by a new parameter $g_s$. This scalar interaction modifies the equations of motion, for which we derive the conserved quantities and the effective potential governing the radial motion of massive particles.

    \item Using this effective potential, we examined the behavior of the specific angular momentum. The analysis reveals that a positive scalar coupling parameter $g_s$ increases the specific angular momentum, whereas negative values of $g_s$ lead to its decrease. The effect of the parameter $q$, analyzed for fixed positive values of $g_s$, is more subtle: it increases the angular momentum for negative values of the angular momentum and decreases it for positive values.

    \item A central part of the investigation performed is focused on the innermost stable circular orbit (ISCO) and the radiative efficiency of accretion, which are key observables. We found that increasing the scalar coupling $g_s$ shifts the ISCO radius inward, bringing it closer to the naked JNW singularity and, consequently, leading to a significant increase in accretion efficiency. The parameter $q$ further modulates this behavior. For positive $g_s$, increasing $q$ also pushes the ISCO upwards and improves efficiency. However, for negative values of $g_s$, the effects of $q$ are reversed, highlighting the complex interplay between spacetime and interaction parameters.

    \item We have also investigated the degeneracy between the parameters of the scalar-tensor model ($q$, $n$, $g_s$) studied and the spin parameter $a$ of the Kerr black hole. By comparing the ISCO radii predicted by the solution obtained with those of the Kerr metric, we demonstrated that certain combinations of parameters investigated can effectively mimic the observational signatures of black hole spin. In particular, small values of the spacetime parameter $n$ produce ISCO shifts comparable to those of rapidly rotating Kerr black holes, even in the absence of intrinsic angular momentum. This degeneracy highlights the importance of multi-frequency QPO modeling and the need for complementary observational constraints to break such ambiguities when testing alternative gravity theories with current and future X-ray observations.

    \item We extended the dynamical study developed to include oscillatory motion by deriving the epicyclic frequencies in the presence of the scalar interaction. The parameter $n$, which characterizes spacetime, was found to have the most dominant effect, causing a decrease in both orbital (Keplerian) frequency $\nu_\phi$ and radial epicyclic frequency $\nu_r$. In contrast, the parameter $q$ induces a slight increase in both frequencies. The scalar coupling $g_s$ has a distinctive frequency-dependent effect: it increases the radial epicyclic frequency $\nu_r$ while simultaneously decreasing the orbital frequency $\nu_\phi$. These frequencies were then used to explore the epicyclic resonance (ER) model for high-frequency QPOs. We demonstrated how the characteristic upper ($\nu_U$) and lower ($\nu_L$) frequencies of this model are sensitive to all three parameters ($n$, $g_s$, $q$), establishing a direct link between the modified gravity theory studied here and potential observational signatures.

    \item To constrain the model developed here with real astrophysical data, we performed a Markov Chain Monte Carlo (MCMC) analysis using astronomical observations of twin-peak QPOs from two well-known microquasars. This robust statistical method allowed us to estimate the best-fit values for the model parameters. The analysis performed yields black hole masses that are consistent with previous estimates: $M/M_{\odot} = 8.99 \pm 0.55$ for XTE J1550-564 and $M/M_{\odot} = 13.3 \pm 1.0 $ for GRS 1915+105. The inferred spacetime parameter $n$ is approximately $0.6$ for both sources. The corresponding observed radii are $r/M \approx 6.3$ and $6.4$, respectively. Furthermore, the analysis provides the first observational constraints on new parameters studied, with $q$ converging to an approximate value $\approx 3$, and the scalar coupling $g_s$ estimated at $0.45$ and $0.44$ for the two sources.

\end{itemize}

In conclusion, we have presented a well-defined extension of the JNW spacetime, thoroughly analyzed its impact on particle dynamics and accretion efficiency, and, for the first time, placed observational constraints on its parameters using QPO data from observed black hole binaries. The obtained results demonstrate that this class of modified gravity theories leaves distinctive imprints on astrophysical observables, making them testable with current and future observations in the strong gravity regime. The consistency of the MCMC fitted with observed data suggests that such scalar-tensor modifications remain a viable and interesting avenue for exploring gravity in the strong-field regime.

\acknowledgments

This research was funded by the National Natural Science Foundation of China (NSFC) under Grant No. U2541210.

\bibliographystyle{JHEP}
\bibliography{references}

\end{document}